\documentclass[usenatbib]{mn2e}
\usepackage{graphicx,tabularx,amsmath,amssymb,upgreek,wasysym,mathtools,multirow}
\numberwithin{equation}{section}
\title[Simulations of star formation in Ophiuchus, II]{Simulations of star formation in Ophiuchus, II: Multiplicity}
\author[O. Lomax, A. P. Whitworth, D. A. Hubber, D. Stamatellos and S. Walch]{O. Lomax\thanks{E-mail: oliver.lomax@astro.cf.ac.uk}$^1$, A. P. Whitworth$^1$, D. A. Hubber$^{2,3}$, D. Stamatellos$^4$ and S. Walch$^{5}$\\
$^1$School of Physics and Astronomy, Cardiff University, Cardiff CF24 3AA, UK\\
$^2$University Observatory, Ludwig-Maximilians-University Munich, Scheinerstr.1, D-81679 Munich, Germany\\
$^3$Excellence Cluster Universe, Boltzmannstr. 2, D-85748 Garching, Germany\\
$^4$Jeremiah Horrocks Institute, University of Central Lancashire, Preston, Lancashire, PR1 2HE, UK\\
$^5$Physikalisches Institut, Universit{\"a}t zu K{\"o}ln, Z{\"u}lpicher Strasse 77, D-50937 Cologne, Germany}

\begin{document}
\pagerange{\pageref{firstpage}--\pageref{lastpage}} \pubyear{2013}
\maketitle
\label{firstpage}

\begin{abstract} 
Lomax et al. have constructed an ensemble of 60 prestellar cores having masses, sizes, projected shapes, temperatures and non-thermal radial velocity dispersions that match, statistically, the cores in Ophiuchus; and have simulated the evolution of these cores using SPH. Each core has been evolved once with no radiative feedback from stars, once with continuous radiative feedback, and once with episodic radiative feedback. Here we analyse the multiplicity statistics from these simulations. With episodic radiative feedback, (i) the multiplicity frequency is $\sim\!60\%$ higher than in the field; (ii) the multiplicity frequency and the mean semi-major axis both increase with primary mass; (iii) one third of multiple systems are hierarchical systems with more than two components; (iv) in these hierarchical systems the inner pairings typically have separations of a few au and mass ratios concentrated towards unity, whereas the outer pairings have separations of order $100\,{\rm au}$ and a flatter distribution of mass ratios. The binary statistics are compatible with observations of young embedded populations, and -- if wider orbits are disrupted preferentially by external perturbations -- with observations of mature field populations. With no radiative feedback, the results are similar to those from simulations with episodic feedback. With continuous radiative feedback, brown dwarfs are under-produced, the number of multiple systems is too low, and the statistical properties of multiple systems are at variance with observation. This suggests that star formation in Ophiuchus may only be representative of global star formation if accretion onto protostars, and hence radiative feedback, is episodic.
\end{abstract}


\section{Introduction}%

Two of the fundemental goals of star formation theory are to explain the origin of the stellar initial mass function (IMF) \citep[e.g.][]{K01,C03,C05} and to explain the statistics of stellar multiple systems \citep[e.g.][]{RMH10,JHB12}. Numerical simulations \citep[e.g.][]{BVB04,DCB04b,BB05,GWW06,B09c,LWHSW14,B12} deliver distributions of stellar mass and/or multiplicity statistics that mimic the observations reasonably well, but it is not yet clear that these simulations are capturing correctly the physical processes actually at work in nature, or whether -- for example -- many different sequences of physical processes in a simulation could be delivering artificially an approximately log-normal distribution of stellar masses. A further complication is that, whereas the mass of a star is believed to be more or less fixed within $0.1\,{\rm Myr}$ of its birth, many of the parameters of multiple systems may evolve for a lot longer. Indeed a significant fraction of primordial multiple systems is likely to be destroyed completely in a clustered environment \citep[e.g.][]{PGKK09,MK11,PR13}.

The semi-major axes of observed multiple systems span more than six orders of magnitude, from contact binaries with $a\sim10^{-2}\,{\rm au}$, to wide binaries with $a\ga 10^4\,{\rm au}$ \citep[see][and references therein]{DK13}. Higher-order hierarchical multiple systems with three to seven stars are observed \citep[e.g.][]{ET08,T08}. Reproducing such systems in numerical simulations is difficult, both because of the variety of physical processes that might play important roles (e.g. magneto-hydrodynamics, radiative and mechanical feedback, impulsive interactions), and because numerical codes have difficulty handling the large range of spatial scales involved. These difficulties are usually addressed either by performing monolithic simulations of whole molecular clouds that spawn many tens or even hundreds of stars \citep[e.g.][]{B09a,B12,GFABK12,GFBK12,WWBWH13,MKKM14,B14} or by performing multiple  simulations of single dense molecular cores that individually spawn $\lesssim10$ stars \citep[e.g.][]{DCB04b,GW04,GWW04,GWW06,WBWNG09,WNWB10,LWHSW14}.

This paper analyses the numerical simulations of prestellar cores presented in \citet{LWHSW14} (hereafter LWH14). There, Smoothed Particle Hydrodynamics (SPH) has been used to model the evolution of an ensemble of dense molecular cores with initial conditions informed as closely as possible by observations of Ophiuchus \citep{MAN98,ABMP07}. The evolution has been simulated first assuming no radiative feedback, then assuming episodic radiative feedback, and finally assuming continuous radiative feedback. Here, we analyse the multiple systems formed in those simulations. In \S \ref{prev_work}, we review the main features of the simulations presented in LWH14. In \S \ref{stellar_mult} we define measures of stellar multiplicity, and the method used to identify multiple systems in the simulations. In \S \ref{results} we present and evaluate the statistical properties of the multiple systems formed in the simulations. Specifically, we first compare and contrast the results obtained with the different treatments of radiative feedback; then we compare the results with observations, both the limited observations of young embedded populations, and the more extensive and statistically robust observations of mature field populations (which have likely been modified by dynamical processing). In \S \ref{discusion} we summarise our main conclusions. In the Appendix, we explain how we derive population statistics from small samples

\section{The simulations}\label{prev_work}%

\subsection{Initial conditions}\label{SEC:INITCONDS}%

\subsubsection{Bulk properties}%

LWH14 generate the bulk properties of the 60 simulated cores by sampling from a trivariate lognormal distribution. They define a vector of core parameters, $\boldsymbol{x}\equiv(\log(M),\log(R),\log(\sigma_\textsc{nt}))$, where $M$ is the core mass, $R$ is the mean core radius and $\sigma_\textsc{nt}$ is the one-dimensional non-thermal velocity dispersion of the core. Values of $\boldsymbol{x}$ are drawn from the distribution
\begin{eqnarray}\label{obsdist}
P(\boldsymbol{x})\!&\!=\!&\!\frac{1}{(2\uppi)^{3/2}|\boldsymbol{\varSigma}|}\exp\!\left(-\frac{1}{2}(\boldsymbol{x}\!-\!\boldsymbol{\mu})^\mathrm{T}\boldsymbol{\varSigma}^{-1}(\boldsymbol{x}\!-\!\boldsymbol{\mu}) \right)\!,
\end{eqnarray}
where
\begin{eqnarray}
\boldsymbol{\mu}\equiv
  \begin{pmatrix}
    \mu_{_M} \\
    \mu_{_R} \\
    \mu_{_{c}}
  \end{pmatrix}\,,
\end{eqnarray}
and
\begin{eqnarray}
\boldsymbol{\varSigma}\!\equiv\!
  \begin{pmatrix}
    \sigma_{_M}^2 & \rho_{_{M,R}}\,\sigma_{_M}\sigma_{_R} & \rho_{_{M,\sigma_\textsc{nt}}}\,\sigma_{_M}\sigma_{_{\sigma_\textsc{nt}}} \\
    \rho_{_{M,R}}\,\sigma_{_M}\sigma_{_R} & \sigma_{_R}^2 & \rho_{_{R,\sigma_\textsc{nt}}}\,\sigma_{_R}\sigma_{_{\sigma_\textsc{nt}}} \\
    \rho_{_{M,\sigma_\textsc{nt}}}\,\sigma_{_M}\sigma_{_{\sigma_\textsc{nt}}} & \rho_{_{R,\sigma_\textsc{nt}}}\,\sigma_{_R}\sigma_{_{\sigma_\textsc{nt}}} & \sigma_{_{\sigma_\textsc{nt}}}^2
  \end{pmatrix}\!.
\end{eqnarray}
Here $\mu_X$ and $\sigma_X$ are the arithmetic mean and standard deviation of $\log(X)$, and $\rho_{X,Y}$ is the Pearson's correlation coefficient of $\log(X)$ and $\log(Y)$. The values of these terms are given in Table \ref{log_params}. They are derived from dust continuum measurements of cores in Ophiuchus by \citet{MAN98} and spectroscopic line-widths measured by \citet{ABMP07}; for details see LWH14.

\begin{table}
\centering
\caption{Identities, arithmetic means, standard deviations and correlation coefficients for the global parameters of cores in Ophiuchus.}
\begin{tabular}{|r|r|r|}\hline
$M\equiv\log_{_{10}}\!\left(\!\frac{M\;}{{\rm M}_{_\odot}}\!\right)$ & $R\equiv\log_{_{10}}\!\left(\frac{R\,}{\rm au}\right)$ & $\sigma_\textsc{nt}\equiv\log_{_{10}}\!\left(\!\frac{\sigma_\textsc{nt}}{{\rm km}/{\rm s}}\!\right)$ \\
$\mu_{_M}=-0.57\;$ & $\mu_{_R}=3.11\;$ & $\mu_{_{\sigma_\textsc{nt}}}=-0.95\;$ \\
$\sigma_{_M}=0.43\;$ & $\sigma_{_R}=0.27\;$ & $\sigma_{_{\sigma_\textsc{nt}}}=0.20\;$ \\
$\rho_{_{M,R}}=0.61\;$ & $\rho_{_{R,\sigma_\textsc{nt}}}=0.11\;$ & $\rho_{_{\sigma_\textsc{nt},M}}=0.49\;$ \\\hline
\end{tabular}
\label{log_params}
\end{table}

\subsubsection{Shapes}%

LWH14 assign intrinsic core shapes to the 60 simulated cores by applying the fitting method described by \citet[][hereafter LWC13]{LWC13}. LWC13 show that the projected shapes of Ophiuchus cores are not signifcantly correlated with their masses, areas or velocity dispersions, and are well fitted by randomly orientated ellipsoids having relative axes given by
\begin{eqnarray}\nonumber
A&=&1\,,\\\label{EQN:M1}
B&=&\exp(\tau\mathcal{G}_\textsc{b})\,,\\\nonumber
C&=&\exp(\tau\mathcal{G}_\textsc{c})\,,
\end{eqnarray}
with $\tau=0.6$. Here, $\mathcal{G}_\textsc{b}$ and $\mathcal{G}_\textsc{c}$ are random deviates drawn from a Gaussian distribution with zero mean and unit standard deviation. Once the axes have been generated using Eqn. (\ref{EQN:M1}), they are reordered so that $A\geq B\geq C$, and normalised so that $A=1$. Absolute values of the core axes are then set to
\begin{eqnarray}\nonumber
A_\textsc{core}&=&\frac{R}{(BC)^{1/3}}\,,\\
B_\textsc{core}&=&BA_\textsc{core}\,,\\\nonumber
C_\textsc{core}&=&CA_\textsc{core}\,.
\end{eqnarray}

\subsubsection{Density profile}%

Observations suggest \citep[e.g.][]{ALL01,HWL01,KWA05,LMR08,RAP13} that, even if they are not in hydrostatic equilibrium, prestellar cores often approximate well to the density profile of a critical Bonnor-Ebert Sphere \citep{E55,B56}, i.e. $\rho=\rho_{_{\rm C}}{\rm e}^{-\psi(\xi)}$, where $\rho_{_{\rm C}}$ is the central density, $\psi$ is the Isothermal Function, $\xi$ is the dimensionless radius, and the boundary is at $\xi_{_{\rm B}}=6.451$ \citep{CW49}. The FWHM of the column-density through a critical Bonnor-Ebert Sphere corresponds to $\xi_{_{\rm FWHM}}=2.424$, so the density at $(x,y,z)$ is given by
\begin{eqnarray}
\xi&=&\xi_{_{\rm FWHM}}\left(\frac{x^2}{A_\textsc{core}^2}+\frac{y^2}{B_\textsc{core}^2}+\frac{z^2}{C_\textsc{core}^2}\right)^{1/2}\,,\\
\rho&=&\frac{M_\textsc{core}\xi_{_{\rm B}}{\rm e}^{-\psi(\xi)}}{4\pi A_\textsc{core}B_\textsc{core}C_\textsc{core}\,\psi'(\xi_{_{\rm B}})}\,,\hspace{0.5cm}\xi<\xi_{_{\rm B}}\,,
\end{eqnarray}
where $\psi'$ is the first derivative of $\psi$.

\subsubsection{Velocity fields}%

LWH14 generate the initial velocity field in a core by first constructing a random Gaussian turbulent field with three-dimensional power spectrum $P_k\!\propto\! k^{-4},\;1\!\leq\!k\!\leq\!64$. The $k\!=\!1$ modes (i.e. the wavelengths corresponding to the size of the core) are then doctored so that they deliver large-scale motions that are centred on the core, \emph{viz.}
\begin{eqnarray}\nonumber
\begin{bmatrix}
  \boldsymbol{a}(1,0,0) \\
  \boldsymbol{a}(0,1,0) \\
  \boldsymbol{a}(0,0,1)
\end{bmatrix}&=&
\begin{bmatrix}
  r_x & \omega_z & -\omega_y \\
  -\omega_z & r_y & \omega_x \\
  \omega_y & -\omega_x & r_z
\end{bmatrix}\,,\\
&&\\\nonumber
\begin{bmatrix}
  \boldsymbol{\varphi}(1,0,0) \\
  \boldsymbol{\varphi}(0,1,0) \\
  \boldsymbol{\varphi}(0,0,1)
\end{bmatrix}&=&
\begin{bmatrix}
  \uppi/2 & \uppi/2 & \uppi/2 \\
  \uppi/2 & \uppi/2 & \uppi/2 \\
  \uppi/2 & \uppi/2 & \uppi/2
\end{bmatrix}\,;
\end{eqnarray}
this is because these modes are presumed to be the motions that have just created the core, or are in the process of dispersing it following an adiabatic bounce. Here $\boldsymbol{a}(\boldsymbol{k})$ and $\boldsymbol{\varphi}(\boldsymbol{k})$ are respectively the amplitudes and phases of mode $\boldsymbol{k}$; and $r_x$, $r_y$, $r_z$, $\omega_x$, $\omega_y$ and $\omega_z$ are random deviates drawn from a Gaussian distribution with zero mean and unit variance. The $r$-terms determine the magnitude of inward (negative) or outward (positive) radial excursions along each axis; and the $\omega$-terms determine the magnitude of rotation about each axis. The overall velocity dispersion of the field (i.e. incorporating all the higher-$k$ modes) is then scaled to a value of $\sigma_\textsc{nt}$ drawn from Eqn. (\ref{obsdist}).

\subsection{Numerical method}%

\subsubsection{Smoothed particle hydrodynamics}%

The LWH14 simulations use the \textsc{seren} grad-$h$ SPH code \citep{HBMW11}. Gravitational forces are computed using a tree, and the \citet{MM97} formulation of time dependent artificial viscosity is invoked. Gravitationally bound regions with density higher than $\rho_\textsc{sink}=10^{-9}\,\mathrm{g}\,\mathrm{cm}^{-3}$ are replaced with sink particles \citep{HWW13}. The equation of state and the energy equation are treated with the algorithm described in \citet{SWBG07}. In all simulations, the SPH particles have mass $m_\textsc{sph}=10^{-5}\,\mathrm{M}_{\odot}$, so the opacity limit ($\ga\!3\times10^{-3}\,\mathrm{M}_{\odot}$) is resolved with $\ga\!300$ particles; an SPH particle typically has $\sim 57$ neighbours. Sink particles have radius $r_\textsc{sink}\simeq0.2\,\mathrm{au}$, corresponding to the smoothing length of an SPH particle with density equal to $\rho_\textsc{sink}$, and the gravitational field of a sink is also softened on this scale. Hence $r_{_{\rm SINK}}$ is in effect the minimum spatial resolution.

\subsubsection{Accretion luminosity}%

In LWH14, each core constructed according to the procedures in \S \ref{SEC:INITCONDS} is evolved three times, first with continuous accretion onto stars but no radiative feedback from stars (hereafter NRF), second with continuous accretion onto stars and hence continuous radiative feedback (CRF), and third with episodic accretion onto stars and hence episodic radiative feedback from stars (ERF). The simulations with no radiative feedback are unrealistic, but provide a point of comparison for the two other feedback modes.

In the CRF simulations, three quarters of the accretion energy is converted into radiation \citep{OKMK09}, i.e.
\begin{eqnarray}\label{star_lum}
L_\star&\simeq&\frac{3GM_\star\dot{M}_\star}{4R_\star}\,;
\end{eqnarray}
for simplicity, LWH14 set the radius of a protostar to be independent of its mass, $R_\star\!\sim\!3\,\mathrm{R}_{\odot}$ \citep{PS93}. CRF has two important consequences. First, circumstellar accretion discs are usually too hot to fragment, and the simulations therefore spawn very few low-mass hydrogen-burning stars or brown dwarfs \citep{K06,B09b,OKMK09,KCKM10,O10,UME10,LWHSW14}. Second, the predicted luminosities are about an order of magnitude higher than those observed in low-mass star forming regions; this is the \emph{luminosity problem} first noted by \citet{KHSS90}.

In the ERF simulations, LWH14 treat accretion using the phenomenological prescription formulated in \citet{SWH11}. This is based on the disc evolution model of \citet{ZHG09,ZHG10} and \citet{Z10}, in which gravitational torques transport angular momentum in the outer parts of a disc, but the inner parts are too warm to admit the gravitational instabilities that moderate these torques. Consequently matter piles up in the inner disc until it becomes so hot that the level of thermal ionisation is sufficient to couple the matter to the magnetic field and thereby support the magneto-rotational instability (MRI). The MRI then effects rapid transport of angular momentum in the inner disc, and the accumulated matter is dumped onto the star, giving rise to a powerful but short-lived burst of luminosity, typically $L_{_\star}\!\sim\!100\,-\,300\,{\rm L}_{_\odot}$. After this the accretion shuts off again, the luminosity falls to $L_{_\star}\!\sim\!0.1\,-\,0.3\,{\rm L}_{_\odot}$, and the disc can cool back down. The period between outbursts is $\sim\!0.01\,\mathrm{Myr}$, which is consistent with observations \citep[e.g][]{SFW13} and provides a window of opportunity for the accretion disc to fragment into low mass stars and brown dwarfs \citep{SWH11,SWH12,LWHSW14}.

\begin{figure}
\centering
\includegraphics[width=\columnwidth]{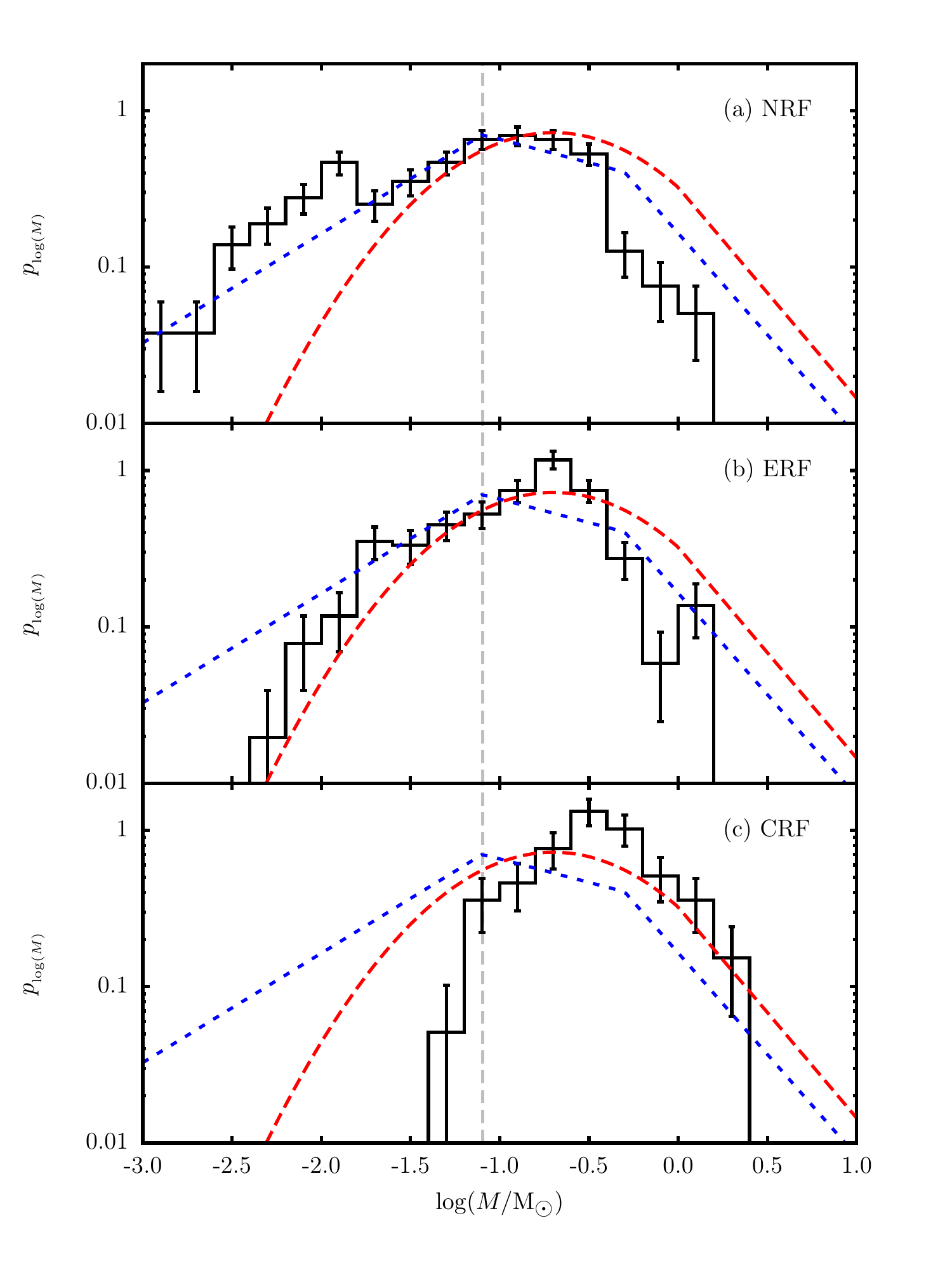}
\caption{The black histograms show stellar mass functions for (a) NRF, (b) ERF, and (c) CRF, with bins of width $\Delta\!\log_{_{10}}\!(M)\!=\!0.2$. The blue dotted straight lines, and the red dashed lognormal curve, show, respectively, the \citet{C05} and \citet{K01} fits to the observed IMF. The vertical dashed line shows the hydrogen burning limit at $M=0.08\,\mathrm{M_\odot}$.}
\label{masses1}
\end{figure}

\subsubsection{Termination of simulations}%

The initial freefall times in the cores are $4_{-2}^{+4}\times 10^4\,{\rm yr}$, and the simulations are terminated after $0.2\,{\rm Myr}$, i.e. $\sim\!5$ freefall times. At this stage typically $70\%$ of the mass has been accreted by protostars, and accretion is still ongoing, so strictly speaking the final masses are lower limits. However, the simulations do not include mechanical feedback, which would tend to slow down, or even terminate accretion. A $70\%$ efficiency for converting cores into stars is compatible both with the theoretical estimate of \citet{MM00} (their upper limit), and with the statistical estimate of \citet{HWGW13} (their lower limit).

\subsection{Stellar statistics}%

\subsubsection{Intercomparison of simulations invoking different radiative-feedback prescriptions}%

The simulations with NRF produce ${\cal N}_{_{\rm S/C}}\!=\!6.0^{+4.0}_{-2.0}$ stars per core, while those with ERF produce ${\cal N}_{_{\rm S/C}}\!=\!3.5^{+3.5}_{-2.5}$, and those with CRF produce ${\cal N}_{_{\rm S/C}}\!=\!1.0^{+0.0}_{-0.0}$; here, we have used the $50^{\rm th}$ centile (i.e. the median), and the $25^{\rm th}$ and $75^{\rm th}$ centiles, in order to quantify non-parametrically the variation in ${\cal N}_{_{\rm S/C}}$ between cores. The reason that ${\cal N}_{_{\rm S/C}}$ decreases going from NRF to ERF to CRF is that, as one progresses along this sequence, fewer and fewer low-mass stars are able to form by disc fragmentation. 

The black histograms on Fig. \ref{masses1} show the mass functions obtained in the simulations with different treatments of radiative feedback. The simulations with NRF give a mass function with $\log_{_{10}}\left(M_{_\star}/{\rm M}_{_\odot}\right)\!=\!-1.10^{+0.36}_{-0.60}$, while those with ERF give $\log_{_{10}}\left(M_{_\star}/{\rm M}_{_\odot}\right)\!=\!-0.82^{+0.22}_{-0.40}$, and those with CRF give $\log_{_{10}}\left(M_{_\star}/{\rm M}_{_\odot}\right)\!=\!-0.48^{+0.19}_{-0.24}\,$; again we have given the median and the $25^{\rm th}$ and $75^{\rm th}$ centiles to represent the spread of stellar masses formed. We see that, as we go from NRF to ERF to CRF, the median mass increases, and the logarithmic spread of masses decreases. The reasons for this trend are twofold. First, as we go from NRF to ERF to CRF, fewer low-mass stars are formed by disc fragmentation. Second, because discs are not fragmenting, more of their mass is available to accrete onto the primary protostar at the disc centre, so the primary stars are somewhat more massive.

The simulations with NRF result in a ratio of low-mass stars to brown dwarfs
\begin{eqnarray}
\mathcal{A}=\frac{N(0.08\,\mathrm{M}_{\odot}<M\leq1.0\,\mathrm{M}_{\odot})}{N(0.03\,\mathrm{M}_{\odot}<M\leq0.08\,\mathrm{M}_{\odot})}\;\,=\;\,2.2\pm 0.3\,,
\end{eqnarray}
whereas the simulations with ERF result in ${\cal A}\!=\!3.9\pm 0.6$, and those with CRF result in ${\cal A}\!=\!17\pm 8$. 

Parenthetically, we note that in all the simulations, stellar masses are not correlated with the masses of their parent cores, suggesting that the form of the IMF is not inherited from the core mass function. Instead, cores with higher mass simply tend to spawn more stars. 

\subsubsection{Comparison of simulations with observation}%

\citet{HWGW13} have shown that the binary statistics of field stars favour a mapping from cores to stars in which a core typically spawns $\overline{\cal N}_{_{\rm S/C}}\!=\!4.3\pm 0.4$ stars (where $\pm 0.4$ is the uncertainty, not the spread). The simulations with ERF agree well with this, whereas those with NRF appear to spawn too many stars per core, and those with CRF too few.

The \citet{C05} IMF has $\log_{_{10}}\!\left(M_{_\star}/{\rm M}_{_\odot}\right)\!=\!-0.70^{+0.50}_{-0.50}$. Again, the ERF simulations are closest to reproducing this. The mass function generated with NRF is skewed too much toward low masses, and that generated with CRF is skewed too much toward high masses.

From an analysis of observations of seven young clusters, \citet{AMGA08} estimate the ratio of low-mass stars to brown dwarfs to be $\mathcal{A}=4.3\pm1.6$. Once again, the simulations with ERF reproduce this result rather well, while those with NRF appear to form too high a proportion of brown dwarfs, and those with CRF too low a proportion of brown dwarfs.

\section{Identifying and characterising multiples}\label{stellar_mult}%

\subsection{System isolation}%

At the end of a simulation, we identify binary systems by interrogating each pair of stars, $(i,j)$, in order of increasing internal energy,
\begin{eqnarray}
E_{ij}&=&M_iM_j\!\left\{\frac{|{\bf v}_i-{\bf v}_j|^2}{2(M_i+M_j)}-\frac{G}{|{\bf r}_i-{\bf r}_j|}\right\}
\end{eqnarray}
 (i.e. the kinetic energy in the centre of mass frame plus the mutual gravitational energy). A pair is identified as a binary system if (i) the internal energy of the pair has remained negative over its last orbital period, and (ii) the net gravitational force due to all the other stars is at least a factor $\alpha_\textsc{p}\!=\!0.1$ less than the gravitational force between the pair. When a binary system with components $i$ and $j$ has been identified, it is replaced with a single virtual star $k$ having mass $M_k\!=\!M_i+M_j$, position ${\bf r}_k\!=\!(M_i{\bf r}_i+M_j{\bf r}_j)/M_k$, velocity ${\bf v}_k\!=\!(M_i{\bf v}_i+M_j{\bf v}_j)/M_k$, and inner separation $\Delta r_k\!=\!|{\bf r}_i-{\bf r}_j|$. The process is then repeated until no pairs satisfying the above criteria are found.

If component $k$ of a pair $k\ell$ is a virtual star, we are dealing with a candidate triple or higher-order multiple system, and a further condition is imposed to ensure that this pairing is tidally stable in the medium term. Specifically, we require that the tidal force exerted on $k$ by $\ell$ is much less (by a factor $\alpha_{_{\rm P}}\!=\!0.1$) than the force holding $k$ together, i.e.
\begin{eqnarray}
\frac{2M_\ell}{|{\bf r}_\ell-{\bf r}_k|^3}&<&\frac{\alpha_\textsc{p}M_k}{\Delta r_k^3}\,,
\end{eqnarray}
throughout the whole of the last orbit. Additionally, if $\ell$ is also virtual, this condition must be satisfied with the roles of $k$ and $\ell$ swapped. In this way we identify not only binaries but also higher-order multiple systems that are hierarchical and therefore unlikely simply to be transient associations.

\subsection{Intrinsic properties of a multiple system}%

A simple binary system, comprising two stars, is characterised by the mass, $M_{_1}$, of the primary (the more massive star); the mass, $M_{_2}$, of the secondary; the mass ratio, $q\!=\!M_{_2}/M_{_1}$; the semi-major axis, $a$, of the orbit; the eccentricity, $e$, of the orbit; the period, $P$, of the orbit; the angles, $\Delta\theta_{_{\rm LS1}},\;\Delta\theta_{_{\rm LS2}}$ between the spins, ${\bf S}_{_1},\;{\bf S}_{_2}$ of the constituent stars and their mutual orbital angular momentum, ${\bf L}_{_{1,2}}$; and the angle $\Delta\theta_{_{\rm S1S2}}$ between the two spins.

An hierarchical multiple of order ${\cal N}$ involves ${\cal N}-1$ orbits. All orbits on which both components are just stars are henceforth referred to as {\sc inner} orbits, and all orbits on which at least one component is a virtual star are henceforth referred to as {\sc outer} orbits. Each orbit is then characterised in essentially the same way as a simple binary system, except that for a virtual star we replace the spin angular momentum with its total angular momentum.

\subsection{Collective properties of multiple systems}%

We adopt the descriptors defined by \citet{RZ93}. These can be divided into those that pertain to stellar systems, and those that pertain to stars.

\subsubsection{System descriptors}%

For a given population, the fraction of stellar systems that is multiple is given by the \emph{multiplicity frequency}
\begin{equation}
mf=\frac{B+T+Q+\ldots}{S+B+T+Q+\ldots}\,,
\end{equation}
where $S,\;B,\;T,\;{\rm and}\;Q$ are the numbers of -- respectively -- single, binary, triple, and quadruple systems\footnote{Strictly speaking, this is the \emph{sample} multiplicity frequency. The underlying \emph{population} multiplicity frequency can be estimated using a Bayesian approach (see Appendix \ref{SEC:MF_APP}).}. The mean number of orbits per system is given by the \emph{pairing factor}
\begin{equation}
pf=\frac{B+2T+3Q+\ldots}{S+B+T+Q+\ldots}\,.
\end{equation}
The mean order of the multiple systems is given by
\begin{equation}
\bar{\mathcal O}_{_{\rm SYS}}=\frac{2B+3T+4Q+\ldots}{B+T+Q+\ldots}\,.
\end{equation}
These system descriptors are commonly estimated for an ensemble of systems in which the primary falls in a restricted interval, and usually this interval can be translated into a range of primary mass\footnote{From a theoretical perspective it might seem more useful to know these system descriptors as a function of {\it system mass} \citep[see][]{G13}, but from a practical observational viewpoint the nature of the primary is the more robustly constrained quantity.}.

\subsubsection{Star descriptors}%

If a star is chosen randomly from a given population, the probability of its being part of a multiple system is given by the \emph{companion probability},
\begin{equation}
cp=\frac{(B_{_1}+B_{_2})+(T_{_1}+T_{_2}+T_{_3})+\ldots}{S+(B_{_1}+B_{_2})+(T_{_1}+T_{_2}+T_{_3})+\ldots}\,,
\end{equation} 
where $S$ is the number of single stars, $B_{_1}$ and $B_{_2}$ are the numbers of -- respectively -- primaries and secondaries in binaries, $T_{_1}$, $T_{_2}$ and $T_{_3}$ are the numbers of -- respectively -- primaries, secondaries and tertiaries in triples, and so on. The mean number of companions per star is given by the \emph{companion frequency},
\begin{equation}
cf=\frac{(B_{_1}+B_{_2})+2(T_{_1}+T_{_2}+T_{_3})+\ldots}{S+(B_{_1}+B_{_2})+2(T_{_1}+T_{_2}+T_{_3})+\ldots}\,.
\end{equation}
If a star is randomly selected from a multiple system within the population, the mean order of its system will be
\begin{equation}
\bar{\mathcal O}_\star=\frac{2(B_{_1}+B_{_2})+3(T_{_1}+T_{_2}+T_{_3})+\ldots}{(B_{_1}+B_{_2})+(T_{_1}+T_{_2}+T_{_3})+\ldots}\,.
\end{equation}
Again, these star descriptors are commonly estimated for an ensemble of stars that fall in a restricted interval, and usually this interval can be translated into a range of stellar mass.

\begin{table*}
  \centering
  \caption{Multiple systems formed in the NRF simulations. Column 1 gives the core ID from LWH14. Column 2 gives the order of the system. Column 3 gives the masses of stars in the system. Here, a binary system is represented by two masses separated by a dash. High order multiples are represented by recursively replacing one or both of the masses with another binary system. Column 4 gives the orbital semimajor axes. Here, a binary system is represented by two $\star$'s separated by a distance value. High order systems are given by replacing one or both $\star$'s with another binary system. Column 5 gives the orbital eccentricities, following the same format as Column 4\,. The full table is given in the online material}
  \begin{tabular}{ccccc}
    \hline
    Core ID & $\mathcal{N}$ & $M$ $(\mathrm{M}_\odot)$ & $a$ $(\mathrm{au})$ & $e$ \\
    \hline
    002 & 2 & (0.062--0.072) & ($\star$--0.2--$\star$) & ($\star$--0.50--$\star$) \\
    002 & 4 & ((0.117--0.229)--(0.038--0.078)) & (($\star$--0.8--$\star$)--21.9--($\star$--0.2--$\star$)) & (($\star$--0.48--$\star$)--0.54--($\star$--0.33--$\star$)) \\
    003 & 4 & ((0.394--0.686)--(0.243--0.347)) & (($\star$--18.0--$\star$)--411.1--($\star$--0.8--$\star$)) & (($\star$--0.29--$\star$)--0.46--($\star$--0.13--$\star$))\\
    \hline
    \vspace{0.2cm}\\
    \multicolumn{5}{l}{\hspace{-0.2cm}\textbf{Table 3.} As Table \ref{onlinematerial1}, but for the systems formed in the ERF simulations.}
    \vspace{0.1cm}\\
    \hline
    Core ID & $\mathcal{N}$ & $M$ $(\mathrm{M}_\odot)$ & $a$ $(\mathrm{au})$ & $e$ \\
    \hline
    002 & 2 & (0.082--0.062) & ($\star$--2.1--$\star$) & ($\star$--0.92--$\star$) \\
    002 & 2 & (0.091--0.284) & ($\star$--2.4--$\star$) & ($\star$--0.03--$\star$) \\
    003 & 5 & (((0.208--0.222)--(0.087--0.132))--0.645) & ((($\star$--0.3--$\star$)--6.6--($\star$--0.2--$\star$))--65.3--$\star$) & ((($\star$--0.03--$\star$)--0.45--($\star$--0.64--$\star$))--0.21--$\star$) \\
    \hline
    \vspace{0.2cm}\\
    \multicolumn{5}{l}{\hspace{-0.2cm}\textbf{Table 4.} As Table \ref{onlinematerial1}, but for the systems formed in the CRF simulations.}
    \vspace{0.1cm}\\
    \hline
    Core ID & $\mathcal{N}$ & $M$ $(\mathrm{M}_\odot)$ & $a$ $(\mathrm{au})$ & $e$ \\
    \hline
    011 & 2 & (0.246--0.332) & ($\star$--9.2--$\star$) & ($\star$--0.02--$\star$) \\
    014 & 2 & (0.062--0.236) & ($\star$--3.4--$\star$) & ($\star$--0.27--$\star$) \\
    029 & 3 & ((0.414--0.458)--0.611) & (($\star$--2.6--$\star$)--24.5--$\star$) & (($\star$--0.37--$\star$)--0.23--$\star$) \\
    \hline
  \end{tabular}
  \label{onlinematerial1}
\end{table*}

\addtocounter{table}{2}

\begin{table*}
\centering
\caption{Population multiplicity frequencies, $mf$, and pairing factors, $pf$, for binary systems having primary masses in the intervals $(0.06,0.1){\rm M}_{_\odot}$ (i.e. very low-mass H-burning stars and brown dwarfs), $(0.1,0.5){\rm M}_{_\odot}$ (i.e. mainly M dwarfs), and $(0.7,1.3){\rm M}_{_\odot}$ (i.e. mainly G dwarfs, aka Sun-like). Row 1 gives the mass range. Rows 2 \& 3 give means and standard deviations of $mf$ and $pf$ for field stars, as estimated by \citet{DK13}. Rows 4 \& 5, 6 \& 7 and 8 \& 9 give means and standard deviations of $mf$ and $pf$ for the simulations with -- respectively -- NRF, ERF and CRF; in addition, the numbers in brackets give the ratios of the simulation means to the observed field means. See the Appendix for how the simulation means and standard deviations are derived.}
\begin{tabular}{|r|l|l|l|}\hline
$M_\star/{\rm M}_{_\odot}$ \hspace{0.8cm} & $0.06\;{\rm to}\;0.1$ & $0.1\;{\rm to}\;0.5$ & $0.7\;{\rm to}\;1.3$ \\
\hline
{\sc Field} \hspace{0.61cm} $mf$ & $0.22\pm0.05$ & $0.26\pm0.03$ & $0.44\pm0.02$ \\
$pf$ & $0.22\pm0.05$ & $0.33\pm0.05$ & $0.62\pm0.03$ \\\hline
NRF \hspace{0.71cm} $mf$ & $0.18\pm0.06\;\;\;(0.82)$ & $0.43\pm0.05\;\;\;(1.65)$ & $0.63\pm0.16\;\;\;(1.43)$ \\
$pf$ & $0.25\pm0.09\;\;\;(1.12)$ & $0.78\pm0.10\;\;\;(2.36)$ & $1.17\pm0.34\;\;\;(1.89)$ \\\hline
ERF \hspace{0.71cm} $mf$ & $0.14\pm0.07\;\;\;(0.64)$ & $0.37\pm0.05\;\;\;(1.42)$ & $0.67\pm0.15\;\;\;(1.52)$ \\
$pf$ & $0.14\pm0.07\;\;\;(0.64)$ & $0.64\pm0.08\;\;\;(1.94)$ & $1.43\pm0.34\;\;\;(2.31)$ \\\hline
CRF \hspace{0.71cm} $mf$ & $0.11\pm0.10\;\;\;(0.50)$ & $0.20\pm0.06\;\;\;(0.77)$ & $0.11\pm0.07\;\;\;(0.25)$ \\
$pf$ & $0.11\pm0.10\;\;\;(0.50)$ & $0.28\pm0.08\;\;\;(0.85)$ & $0.18\pm0.18\;\;\;(0.29)$ \\\hline
\end{tabular}
\label{binarytable}
\end{table*}

\begin{figure}
\centering
\includegraphics[width=0.8\columnwidth]{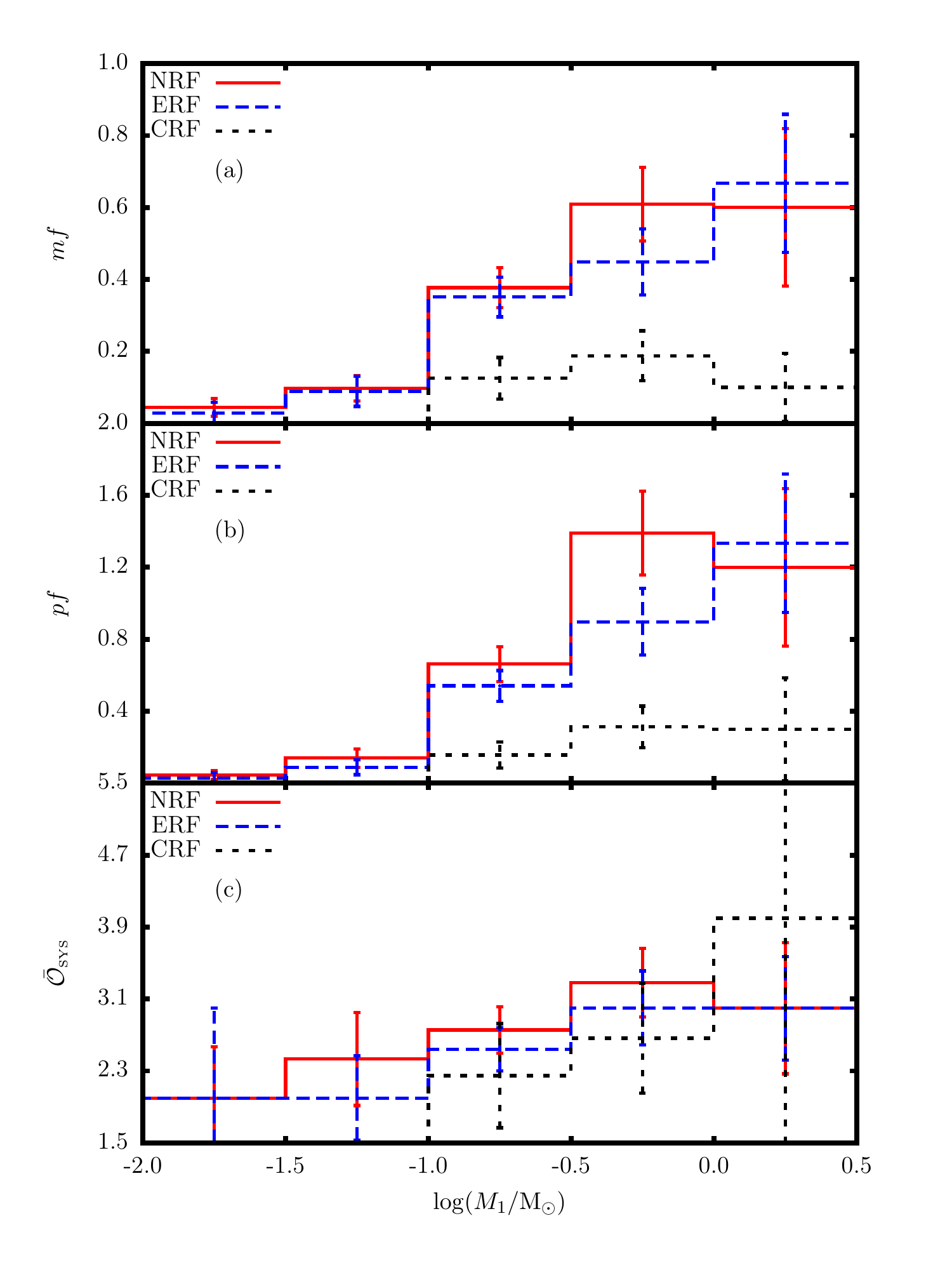}
\caption{(a) Multiplicity frequency, $mf$, (b) pairing factor, $pf$, and (c) mean system order, $\bar{\cal O}_{_{\rm SYS}}$, as a function of primary mass, $M_{_1}$. The solid red histograms give the results obtained with NRF; the blue dashed histograms, ERF; and the dotted black histograms, CRF. Error bars are Poisson counting uncertainties.}
\label{multfreq}
\end{figure}

\begin{figure}
\centering
\includegraphics[width=0.8\columnwidth]{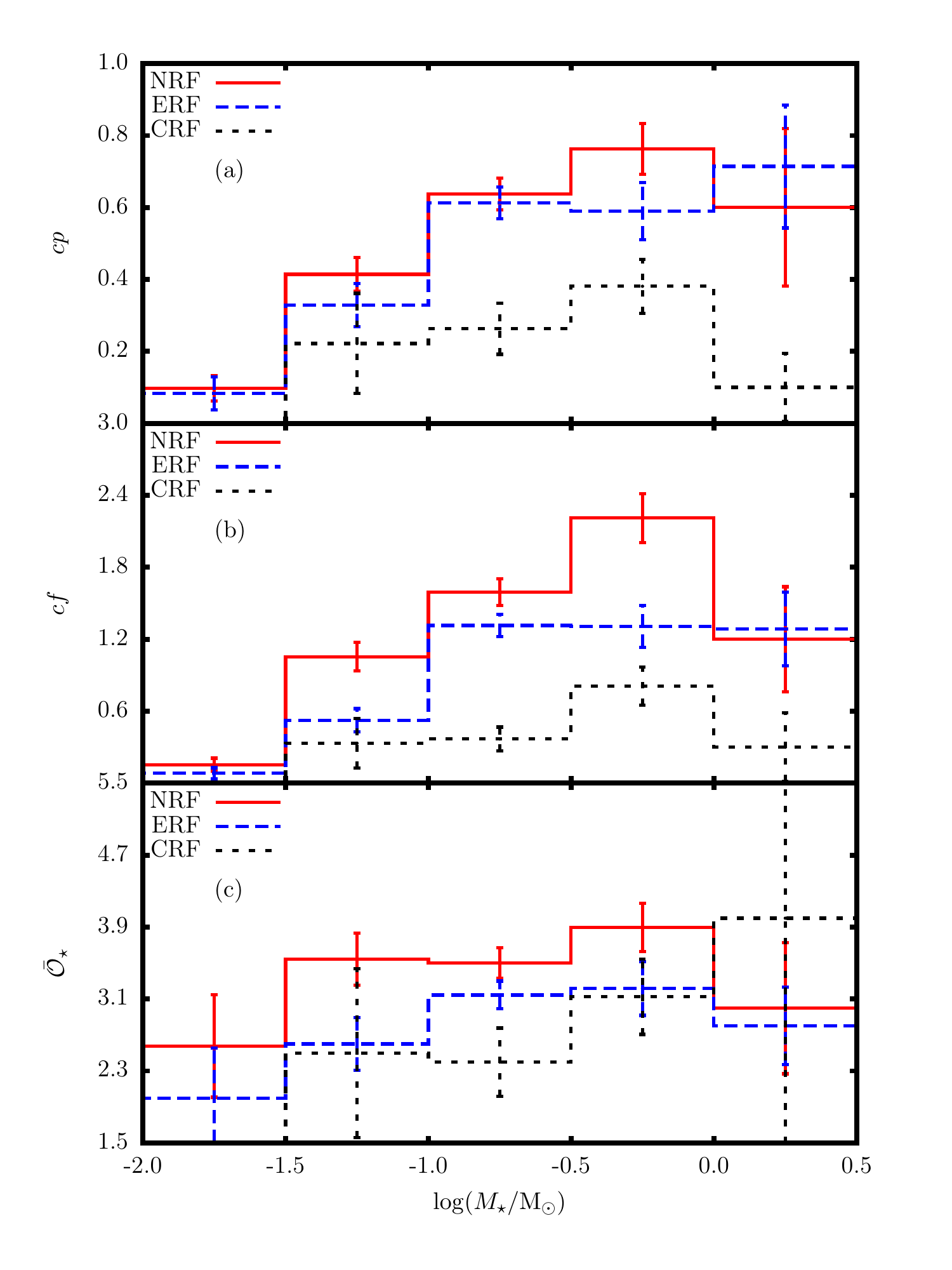}
\caption{(a) Companion probability, $cp$, (b) companion frequency, $cf$, and (c) mean system-order, $\bar{\cal O}_\star$, as a function of stellar mass, $M_\star$. The solid red histograms give the results obtained with NRF; the blue dashed histograms, ERF; and the dotted black histograms, CRF. Error bars are Poisson counting uncertainties.}
\label{comprob}
\end{figure}

\begin{figure}
\centering
\includegraphics[width=0.8\columnwidth]{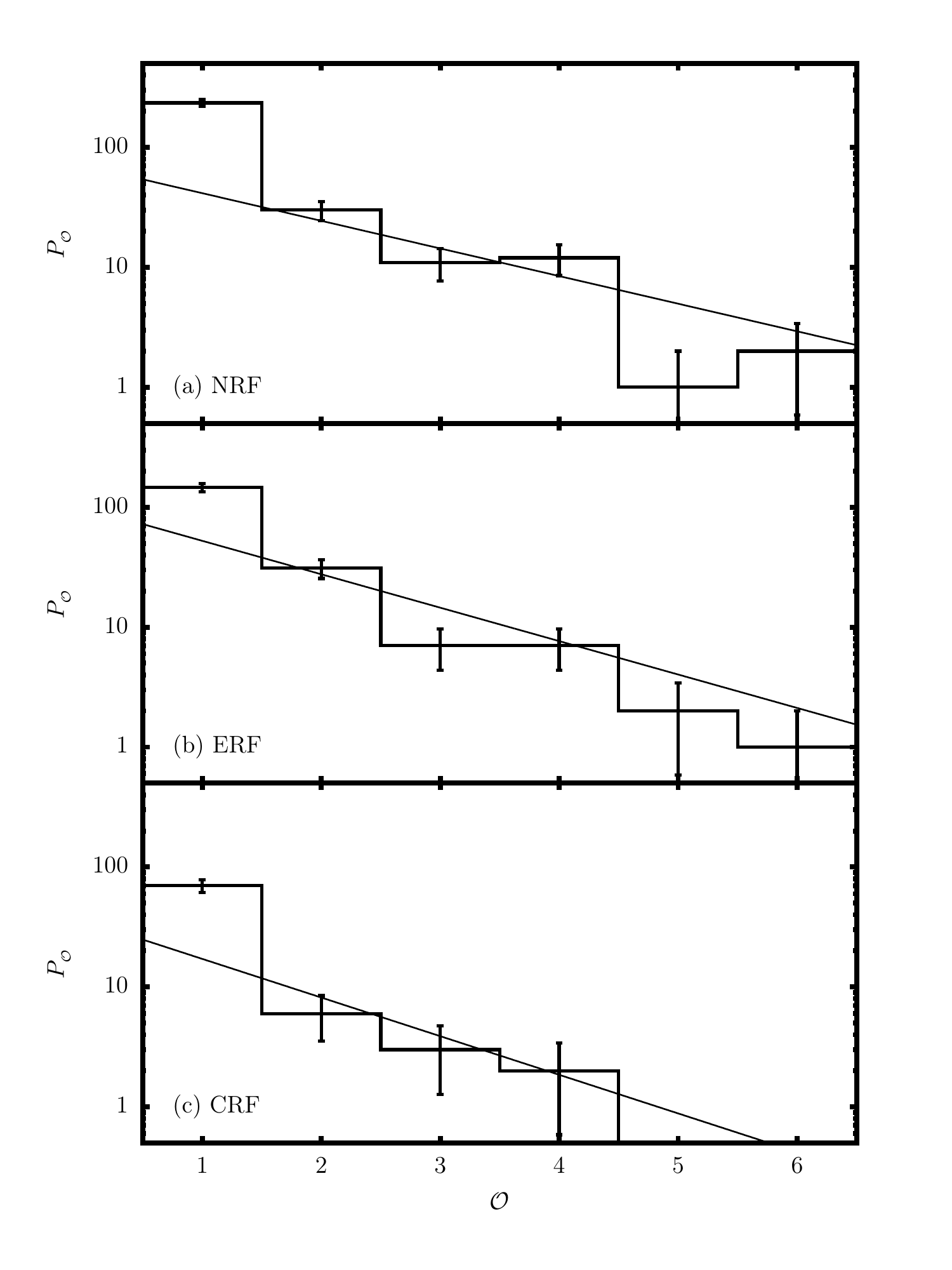}
\caption{The distribution of system orders for simulations with (a) NRF, (b) ERF, and (c) CRF. The histograms show the number of systems of order $\mathcal{O}$ having primary mass greater than $0.1\,\mathrm{M}_\odot$. The thin black lines are best fits to the histograms for ${\cal O}\!\geq\!2$, using $P_{_{\mathcal O}}\!\propto\! {\cal H}^{-\mathcal{O}}$. Values of ${\cal H}$ are given in Table \ref{mult_stats}. Error bars are Poisson counting uncertainties.}
\label{sys_count}
\end{figure}

\section{Results}\label{results}%

Detailed descriptions of the individual multiple systems formed in the simulations are given in Tables \ref{onlinematerial1}, 3 and 4. In this section we analyse their statistical properties, and compare them, firstly with one another, to explore the consequences of the different prescriptions for radiative feedback ({\sc nrf}, {\sc erf} and {\sc crf}), and secondly with observations. In comparing with observations, we distinguish two populations. The population most relevant to the newly-formed stars in our simulations are {\it young embedded stars}, but unfortunately the statistical data on this population is limited to rather weak constraints on the multiplicity frequency, the distribution of higher-order multiples, the distribution of semi-major axes, and the distribution of mass ratios. We therefore also look to the population of {\it mature field stars}, which have almost certainly been "processed" by impulsive and secular interactions with other stars and residual gas before arriving in the field. The advantage of this population is that there exist more robust statistics for it. Furthermore, we have some idea of the systematics of what processing does to a population of multiple stars: specifically, (i) it does not change the distribution of masses significantly; (ii) it increases the number of single stars, both by unbinding wide systems, and through sling-shot ejections from non-hierarchical multiples \citep[e.g.][]{PGKK09}; (iii) it tends to widen the distribution of semi-major axes (equivalently, periods), in particular populating close orbits \citep[e.g.][]{BBB02a}; (iv) it may shift the distribution of mass ratios towards high $q$ (companions of comparable mass) \citep[e.g.][]{KW12}; (v) it probably alters significantly the distribution of orbital eccentricities and the level of spin/orbit and spin/spin alignment \citep[e.g.][]{ST02}. We discuss each of these effects in more detail in the separate sections below, which deal with multiplicity frequency (\S \ref{SEC:MULT}); higher-order multiple systems (\S \ref{SEC:HOM}); semi-major axes (\S \ref{SEC:SMA}); mass ratios (\S \ref{SEC:q}); eccentricities (\S \ref{SEC:ECC}); spin/orbit and spin/spin alignments (\S \ref{SEC:SOA}).

\subsection{Multiplicities}\label{SEC:MULT}%

\subsubsection{Intercomparison of simulations invoking different radiative-feedback prescriptions}%

Figs. \ref{multfreq}a, b and c show the variation of -- respectively -- the multiplicity frequency, $mf$, pairing factor, $pf$, and mean system order, $\cal{\cal O}_{_{\rm SYS}}$, with {\it primary} mass, $M_{_1}$, for the systems formed in the simulations. For NRF and ERF, $mf$ and $pf$ both increase sharply with increasing $M_{_1}$, implying that multiple systems are formed preferentially with at least one massive component. In contrast, for CRF the increase is very modest, implying no strong preference for a massive component; however, this conclusion is moderated by the fact that with CRF there is a smaller range of masses, and the number of multiple systems is lower so the statistics are poorer. For all three radiative feedback prescriptions, $\cal{\cal O}_{_{\rm SYS}}$ increases steadily with $M_{_1}$, implying that higher-order multiples tend to contain at least one relatively massive component.

Figs. \ref{comprob}a, b and c show, respectively, how the companion probability, $cp$, companion frequency, $cf$, and mean system order, $\cal{\cal O}_{_{\rm STAR}}$, vary with {\it stellar} mass, $M_\star$, for the stars formed in the simulations. For all three feedback prescriptions, both $cp$ and $cf$ increase slowly with increasing mass, and then fall off at the highest masses. This fall-off reflects two factors. First, low-mass stars feature more strongly in these statistics, because they count as being in a binary even if they are only the secondary, as being in a triple even if they are only the secondary or tertiary, and so on. Second, the highest mass stars are usually ones that have formed from slowly rotating cores; they have therefore consumed most of the core mass and been attended by low-mass discs, so the capacity for forming companions has been reduced.

\subsubsection{Comparison of simulations with observation}%

{\it Young embedded stars.} 

\noindent Observational estimates of multiplicity frequencies in young embedded populations \citep[e.g.][]{RKL05,LZW93,KIM11} are both too sparse, and too compromised by selection effects, to justify a detailed comparison with the simulation results. However, these estimates do suggest that the multiplicity frequencies of young embedded populations exceed those of mature field populations by a factor between one and two. This factor gives us a rough measure of the fraction of systems that has been disrupted by the time it reaches the field. \\

\noindent {\it Mature field stars.} 

\noindent In Table \ref{binarytable} we compare the observed multiplicity frequencies and pairing factors for field stars given by \citet{DK13} in specific mass bins with their counterparts from the simulations.

At low masses, $0.06\,{\rm M}_{_\odot}\la M_{_1}\la 0.10\,{\rm M}_{_\odot}$, i.e. systems with primaries that are brown dwarfs or very low-mass H-burning stars, the multiplicity frequency from the simulations with NRF is somewhat below that observed in the field, but easily compatible within the uncertainties; the multiplicity frequency from the simulations with ERF is further below that observed in the field, but again still easily compatible within the uncertainties. We conclude that the low-mass multiplicity statistics observed in the field are reproduced  with both NRF and ERF, but there is little room for $mf$ to be reduced further by processing; this would be acceptable, if very low-mass binaries, being very close (see \S \ref{SEC:SMA}), were not easily disrupted \citep{PR13}. With CRF no low-mass multiple systems are formed, but since only seven systems (all singles) are formed in this mass-bin, this result is formally compatible with what is observed in the field; to make a sensible comment on forming low-mass binaries with CRF, it seems that we would need to do many more simulations, but the indications are that CRF is not conducive to the formation of binaries.

At intermediate masses, $0.1\,{\rm M}_{_\odot}\la M_{_1}\la 0.5\,{\rm M}_{_\odot}$, i.e. systems with mid to late M-dwarf primaries, the multiplicity frequency from the simulations with NRF is $1.7$ times that observed in the field; with ERF, it is $1.4$ times that observed in the field; and with CRF, $0.77$ times. The results obtained with NRF and ERF are easily compatible with the expectation that processing reduces the multiplicity frequency by a factor between one and two; the results obtained with CRF are just compatible, if there is very little reduction due to processing.

At high masses, $0.7\,{\rm M}_{_\odot}\la M_{_1}\la 1.3\,{\rm M}_{_\odot}$, i.e. systems with solar-type primaries, the multiplicity frequency from the simulations with NRF is $1.4$ times that observed in the field; with ERF, it is $1.5$ times that observed in the field; and with CRF, $0.25$ times. The results obtained with NRF and ERF are again easily compatible with the expectation that processing reduces the multiplicity frequency by a factor between one and two; those obtained with CRF are not, and although they are compromised by poor statistics, the clear inference from the simulations is that CRF inhibits the formation of low-mass stars and binaries.

\begin{figure}
\centering
\includegraphics[width=0.98\columnwidth]{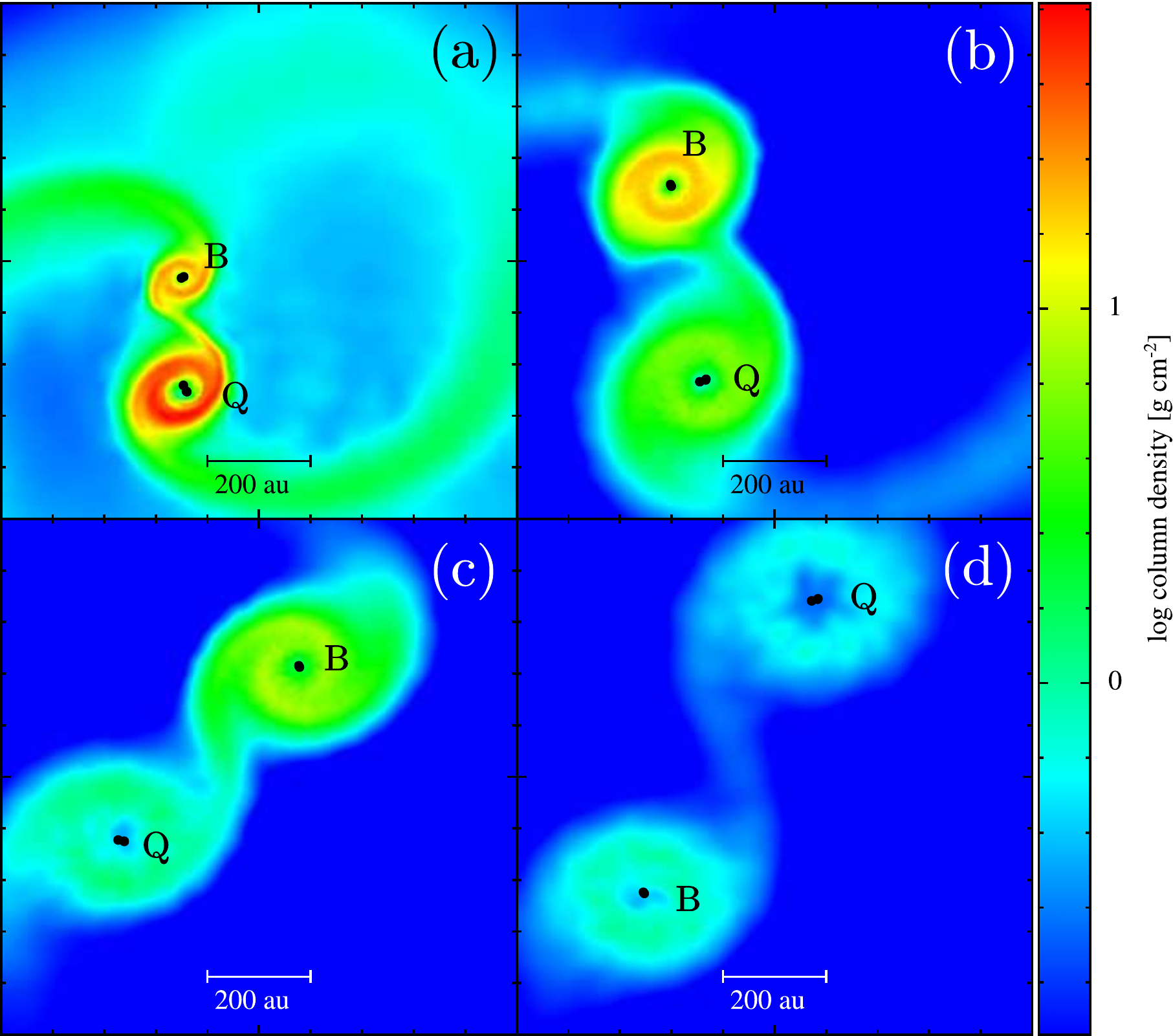}
\caption{False-colour column-density images of the inner $1000\,{\rm au}$ of the core that forms a sextuplet system with ERF, projected on the $y\!=\!0$ plane, at (a) $0.05\,{\rm Myr}$, (b) $0.10\,{\rm Myr}$, (c) $0.15\,{\rm Myr}$, and (d) $0.20\,{\rm Myr}$. The letters {\bf B} and {\bf Q} label the binary and quadruple components of the system. An animation of this figure is given in the online material.}
\label{FIG:SEXT_TIME}
\end{figure}

\begin{figure}
\centering
\includegraphics[width=0.98\columnwidth]{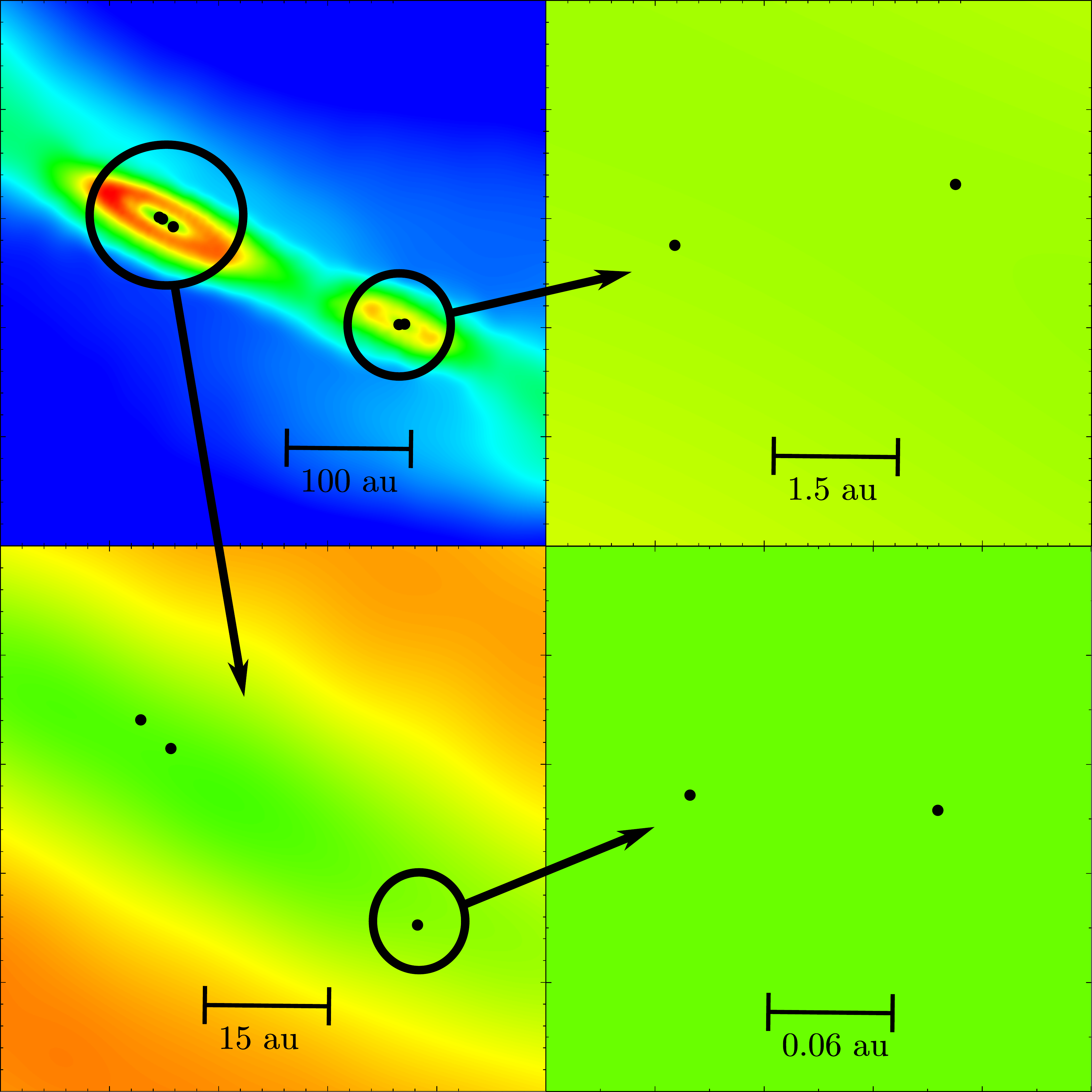}
\caption{A montage of false-colour column-density images of the sextuplet system that formed with ERF, projected on the $z\!=\!0$ plane, at $0.20\,{\rm Myr}$, showing the scales of the different orbits involved, and how they fit together (see also Fig. \ref{FIG:SEXT_PROPS}).}
\label{FIG:SEXT_ZOOM}
\end{figure}

\begin{figure}
\centering
\includegraphics[width=0.98\columnwidth]{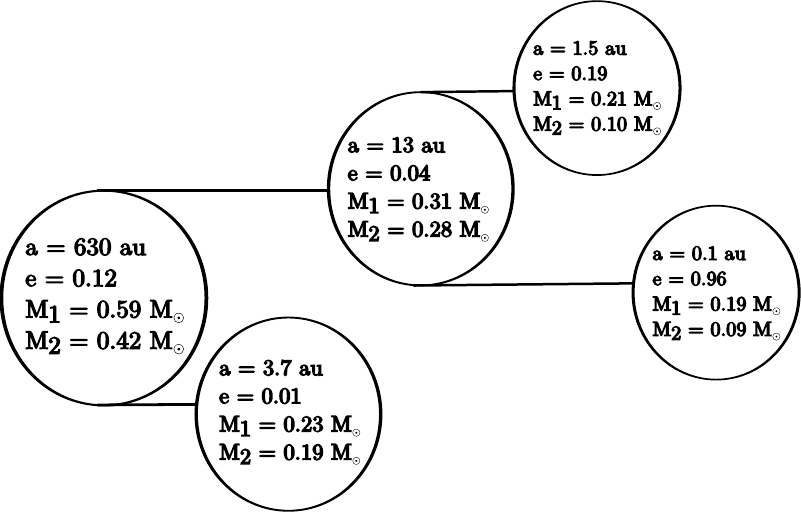}
\caption{A schematic showing the masses and orbital parameters of the different components of the sextuple formed with ERF. Each circle represents an orbit (thus five orbits for an hierarchical sextuplet), and the circle contains the semi-major axis of the orbit, its eccentricity, and the masses of the two components. These may themselves be subsystems; only if there is nothing to the right of a circle are its components single stars (i.e. these circles are the ones that correspond to {\sc inner} orbits).}
\label{FIG:SEXT_PROPS}
\end{figure}

\subsection{Hierarchical higher-order multiple systems}\label{SEC:HOM}%

\subsubsection{Intercomparison of simulations invoking different radiative-feedback prescriptions}%

Fig. \ref{sys_count} shows the number of systems of different order, ${\cal O}$, produced by the simulations, viz.
\begin{center}
\begin{tabular}{lccccccc}
${\cal O}=$ & 1 & 2 & 3 & 4 & 5 & 6 & {\sc (No.of}$\;\,$ \\
 & $S$ & $B$ & $T$ & $Q$ & $Q'$ & $S'$ & $\;\;\;\;\;\;${\sc orbits)}$\;\;\;\;\;\;$ \\
 & & & & & & \\
NRF\hspace{0.5cm} & 235 & 30 & 12 & 12 & 1 & 2 & (104) \\
ERF               & 146 & 31 & 7  & 7  & 2 & 1 & (79) \\
CRF               & 70  & 6  & 3  & 2  & 0 & 0 & (18) \\
\end{tabular}
\end{center}
Here, $Q'$ and $S'$ are the numbers of quintuples and sextuples, and the final column (in brackets) gives the total number of orbits, viz. $\,B\!+\!2T\!+\!3Q\!+\!4Q'\!+\!5S'$.

Also plotted on Fig. \ref{sys_count} is a geometric fit of the form $P_{_{\mathcal O}}\!\propto\! {\cal H}^{-\mathcal{O}}$, where $\mathcal{O}$ is the multiplicity order of a system (${\cal O}\!=\!1$ for a single, ${\cal O}\!=\!2$ for a binary, etc.), $P_{_{\mathcal O}}$ its relative probability, ${\cal H}$ is the base of the distribution, and the fit is made to all systems with $\mathcal{O}\!\geq\!2$. The lower ${\cal H}$ is, the higher is the proportion of high-order multiple systems relative to binaries. All three sets of simulations give ${\cal H}\!\sim\!2$ (see Table \ref{mult_stats}), which implies that the simulations deliver quite a high proportion of high-order multiples. Specifically, $\,\sim\!70\%$ of all stars with $M_{\star}>0.1\,\mathrm{M}_\odot$ are in multiple systems, and the mean order of these multiple systems is $\sim\!3$, meaning that roughly half are triples or higher-order systems.

\subsubsection{Comparison of simulations with observation}%

{\it Young embedded stars.} 

\noindent Observations of young embedded populations also reveal many high-order multiples including sextuples \citep[e.g.][]{KIM11}, but the statistics are too poor to merit a parametric fit. \\

\noindent {\it Mature field stars.} 

\noindent \citet{ET08} find ${\cal H}\sim3.4$ for Sun-like Main Sequence stars in the field, and \citet{DK13} find a similar value for M-dwarfs in the field. In other words, the proportion of high-order multiples in the field seems to be significantly lower than the proportion we obtain for the much younger systems produced in our simulations. This is consistent with the idea that (the often rather low-mass) stars on wide orbits in high-order multiples are preferentially ionised, either by internal relaxation, or by tidal interactions with other stars and gas clouds.

To illustrate the processes by which high-order multiples form in the simulations, we focus on the core that forms a sextuplet system when evolved with ERF. Fig. \ref{FIG:SEXT_TIME} shows a false-colour column-density image of the inner $1000\,{\rm au}$ of the core at 0.05, 0.10, 0.15 and $0.20\,{\rm Myr}$ (the end of the simulation). Fig. \ref{FIG:SEXT_ZOOM} zooms in on the final frame to dissect the hierarchical nature of this system. Fig. \ref{FIG:SEXT_PROPS} presents schematically the masses and orbital parameters of the different levels of the system. This system spans a huge range of linear scale, from the tightest inner orbit ($a\!\sim\!0.1\,{\rm au}$) to the widest outer orbit ($a\!\sim\!630\,{\rm au}$). We note that the close pairing with $a\!\sim\!0.1\,{\rm au}$ actually forms at a separation of nearly $100\,{\rm au}$, and is then scattered into its final position; therefore its formation is not an artefact of the creation of sink particles. A seventh star is also formed, but it is ejected. The same core evolved with NRF spawns a triple, a binary and a single (six stars in total); when evolved with CRF this core just spawns a binary.

\subsection{Semimajor axes}\label{SEC:SMA}%

\begin{figure}
\centering
\includegraphics[width=0.75\columnwidth]{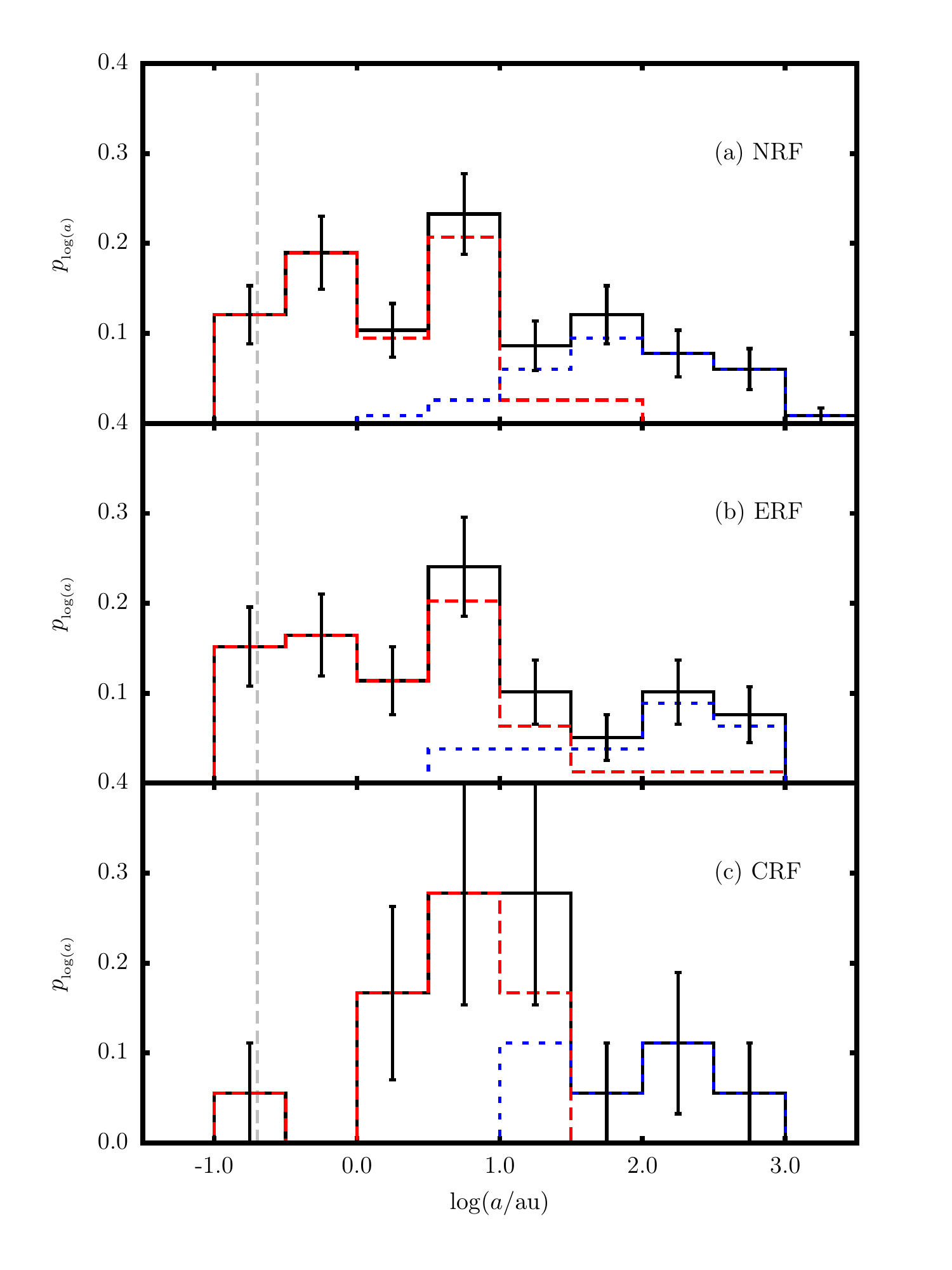}
\caption{The distribution of orbital semimajor axes from the simulations with (a) NRF, (b) ERF, and (c) CRF. The solid back histograms show the distributions for all orbits; the dashed red histograms, for {\sc inner} orbits only; and the dotted blue histograms, for {\sc outer} orbits only. Error bars are derived from Poisson counting uncertainties. The vertical dashed line marks the sink radius, $r_{_{\rm SINK}}\!=\!0.2\,{\rm au}$.}
\label{semimajor}
\end{figure}

\begin{figure}
\centering
\includegraphics[width=0.8\columnwidth]{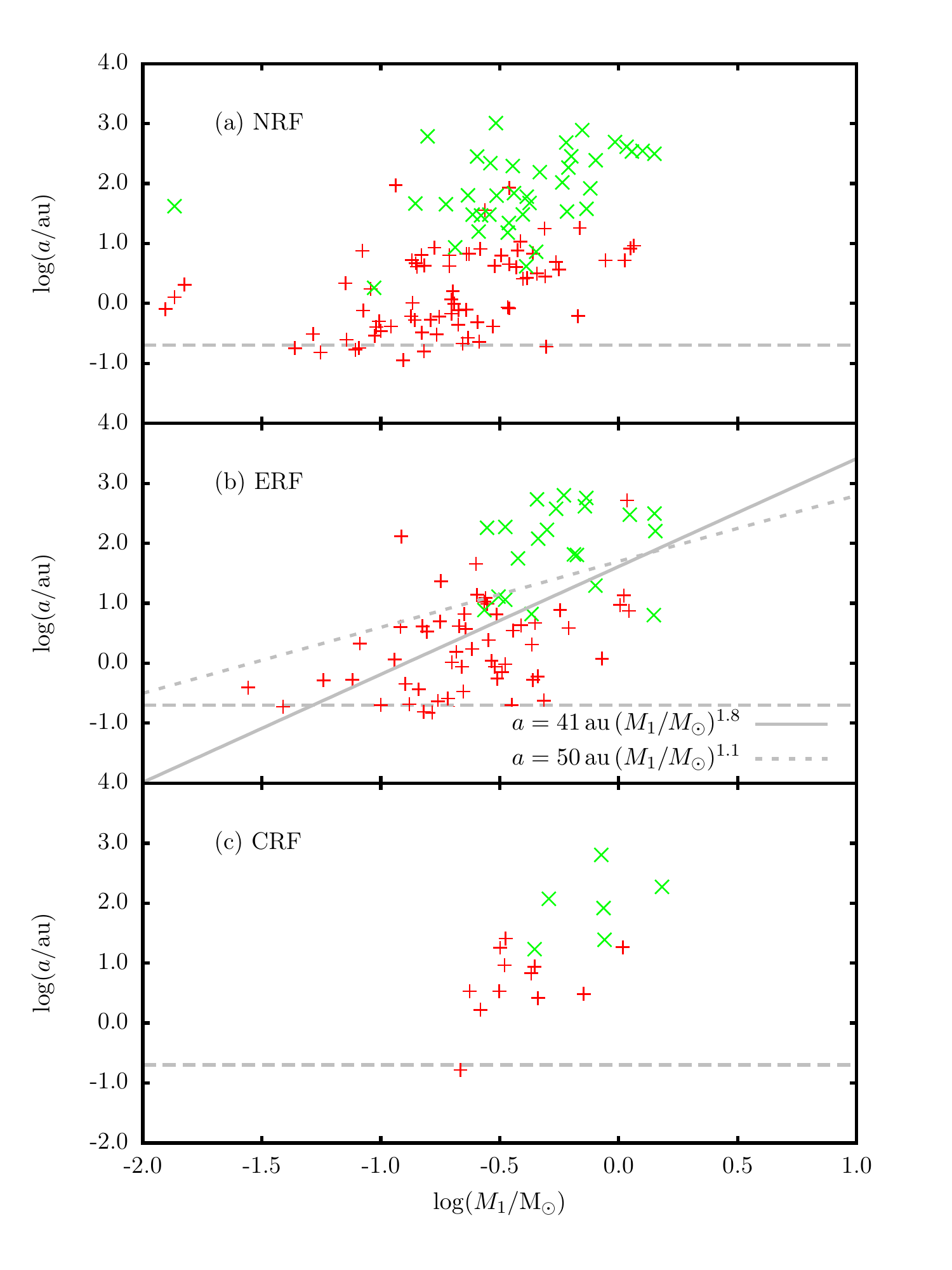}
\caption{Orbital semimajor axis, $a$, plotted against primary mass, $M_{_1}$, for simulations with (a) NRF, (b) ERF, and (c) CRF. A red + represents an {\sc inner} orbit. A green $\times$ represents an {\sc outer} orbit. In frame (b), the dashed line is an approximate power-law fit to the  variation of mean semi-major axis with primary mass, as observed for Sun-like stars and M Dwarfs in the field: ${\bar a}\!\sim\!50\,{\rm au}\,(M_{_1}/{\rm M}_{_\odot})^{1.1}$. The solid line is an approximate power-law fit to the variation of mean semi-major axis with primary mass, for primary masses $M_{_1}\!>\!0.1\,{\rm M}_{_\odot}$, as seen in the simulations with ERF: ${\bar a}\!\sim\!41\,{\rm au}\,(M_{_1}/{\rm M}_{_\odot})^{1.8}$.}
\label{primass_semimajor}
\end{figure}

\begin{table}
\centering
\caption{Multiplicity statistics for the simulations with NRF, ERF and CRF. The first block (rows 2 to 5) gives, for all orbits, the mean, $\mu_a$, and standard deviation, $\sigma_a$, of $\log_{_{10}}(a/{\rm au})$, where $a$ is the semi-major axis; the best-fitting exponent, $\gamma$, for the distribution of mass ratios ($p_q\!\propto\!q^\gamma$); and the best-fitting base, ${\cal H}$, for the distribution of multiplicity orders ($P_{\cal O}\!\propto\!{\cal H}^{-{\cal O}}$). The second block gives $\mu_a$, $\sigma_a$ and $\gamma$ for {\sc inner} orbits only. The third block gives the same information for {\sc outer} orbits only. The fourth block give $\gamma$ for orbits with semi-major axis less than or greater than $5\,{\rm au}$.}
\begin{tabular}{|r|c|c|c|}\hline
$a\!\equiv\!\log_{_{10}}(a/{\rm au})$ & NRF & ERF & CRF \\\hline
{\sc All} \hspace{1.04cm} $\mu_a$ & $0.8\pm0.1$ & $0.7\pm0.1$ & $1.1\pm0.2$ \\
$\sigma_a$ & $1.0\pm0.1$ & $1.1\pm0.1$ & $0.8\pm0.2$ \\
$\gamma$ & $0.6\pm0.1$ & $0.9\pm0.2$ & $1.3\pm0.5$ \\
${\cal H}$ & $1.7^{+0.2}_{-0.2}$ & $1.9^{+0.3}_{-0.2}$ & $2.1^{+0.6}_{-0.5}$ \\\hline
{\sc Inner} \hspace{0.74cm} $\mu_a$ & $0.2\pm0.1$ & $0.2\pm0.1$ & $0.7\pm0.2$ \\
$\sigma_a$ & $0.7\pm0.1$ & $0.8\pm0.1$ & $0.6\pm0.2$ \\
$\gamma$ & $0.7\pm0.2$ & $1.0\pm0.3$ & $2.4\pm0.9$ \\\hline
{\sc Outer} \hspace{0.64cm} $\mu_a$ & $1.9\pm0.1$ & $1.9\pm0.2$ & $2.0\pm0.2$ \\
$\sigma_a$ & $0.9\pm0.1$ & $0.7\pm0.2$ & $0.5\pm0.2$ \\
$\gamma$ & $0.4\pm0.2$ & $0.8\pm0.2$ & $0.6\pm0.6$ \\\hline
$a\!<\!5\,{\rm au}$ \hspace{0.61cm} $\gamma$ & $1.1\pm0.3$ & $1.0\pm0.3$ & $2.3\pm1.1$ \\
$a\!>\!5\,{\rm au}$ \hspace{0.61cm} $\gamma$ & $0.3\pm0.2$ & $0.7\pm0.3$ & $1.2\pm0.6$ \\\hline
\end{tabular}
\label{mult_stats}
\end{table}

\begin{figure}
\centering
\includegraphics[width=0.8\columnwidth]{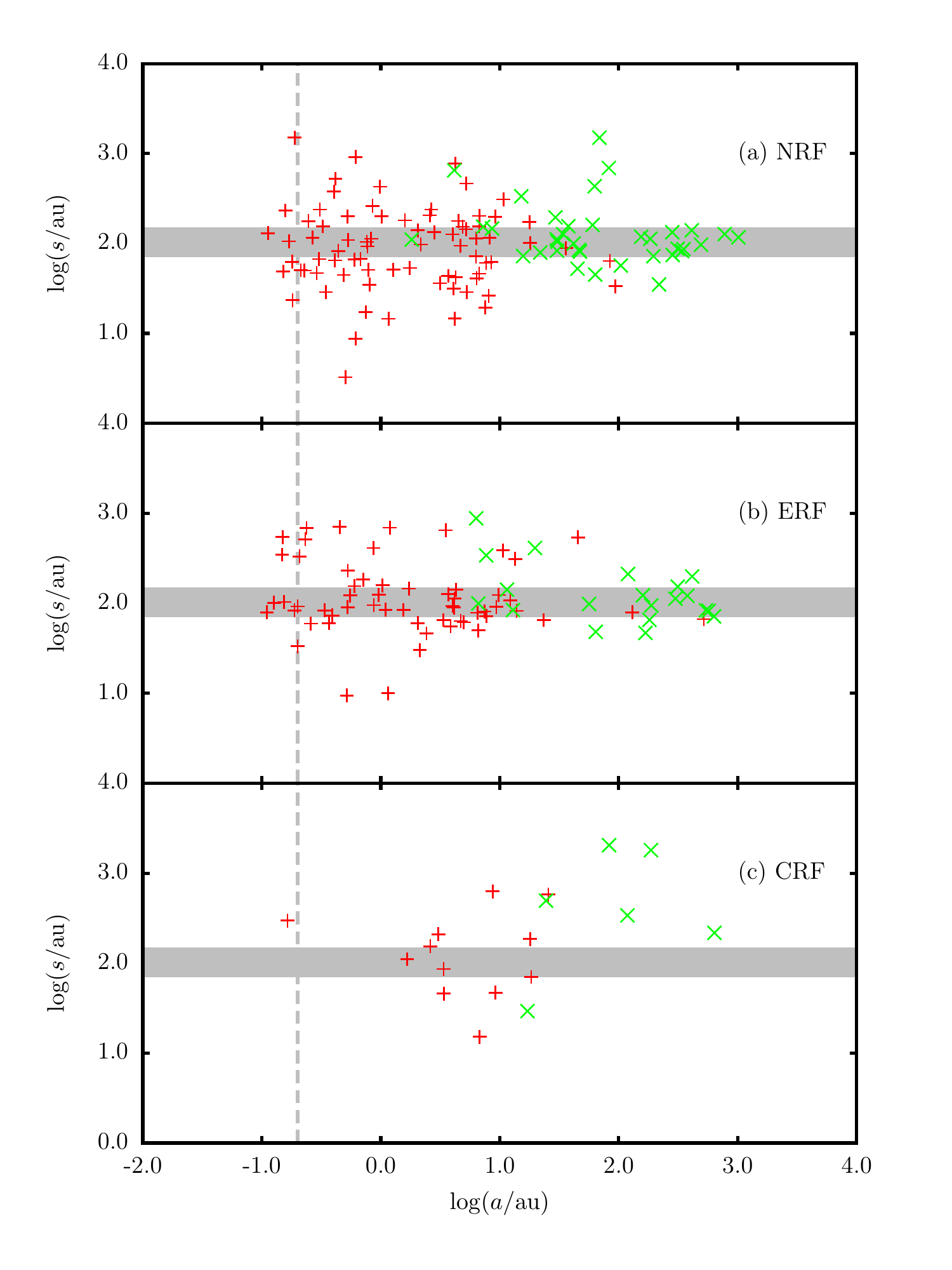}
\caption{The separation between sink particles at their creation, $s$, plotted against orbital semimajor axis, $a$, at $0.2\,{\rm Myr}$ for simulations with (a) NRF, (b) ERF, and (c) CRF. A red + represents an {\sc inner} orbit. A green $\times$ represents an {\sc outer} orbit. The grey horizontal bar marks the range $70\,{\rm au}$ to $140\,{\rm au}$, which is the range of separations favoured by disc fragmentation \citep[see][]{WS06}. The vertical dashed line marks the sink radius, $r_{_{\rm SINK}}\!=\!0.2\,{\rm au}$.}
\label{FIG:s2a}
\end{figure}

\begin{figure}
\centering
\includegraphics[width=0.8\columnwidth]{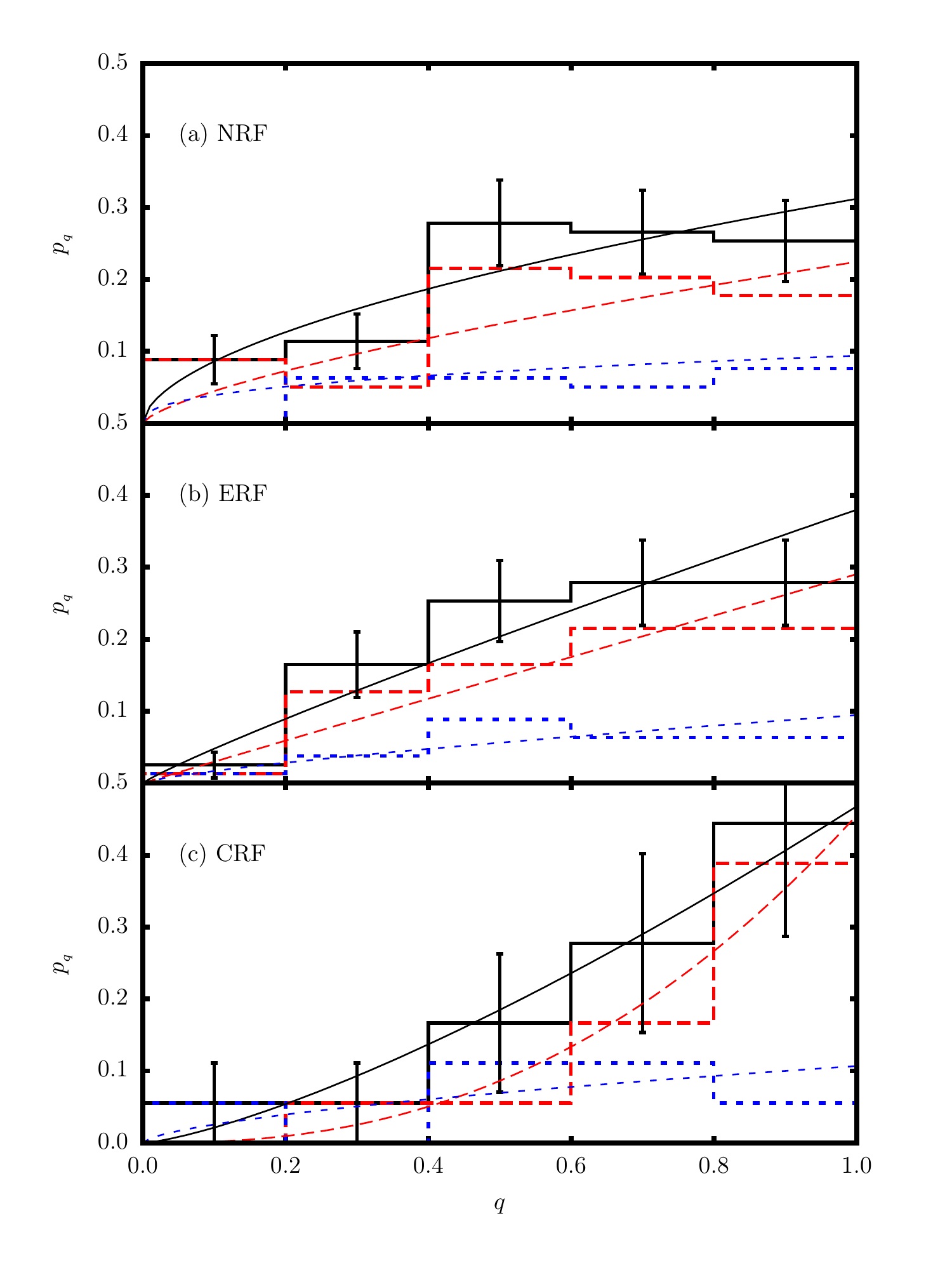}
\caption{The solid back histograms shows the distribution of mass ratios, $q$, for all multiple systems in simulations with (a) NRF, (b) ERF, and (c) CRF. The dashed red histograms show the $q$-distributions for {\sc inner} orbits only. The dotted blue histograms show the $q$-distributions for {\sc outer} orbits only. The over-plotted thin lines are power-law fits of the form $p_q\!\propto\!q^\gamma$; the corresponding values of $\gamma$ are given in Table \ref{mult_stats}.}
\label{massratios}
\end{figure}

\begin{figure}
\centering
\includegraphics[width=0.8\columnwidth]{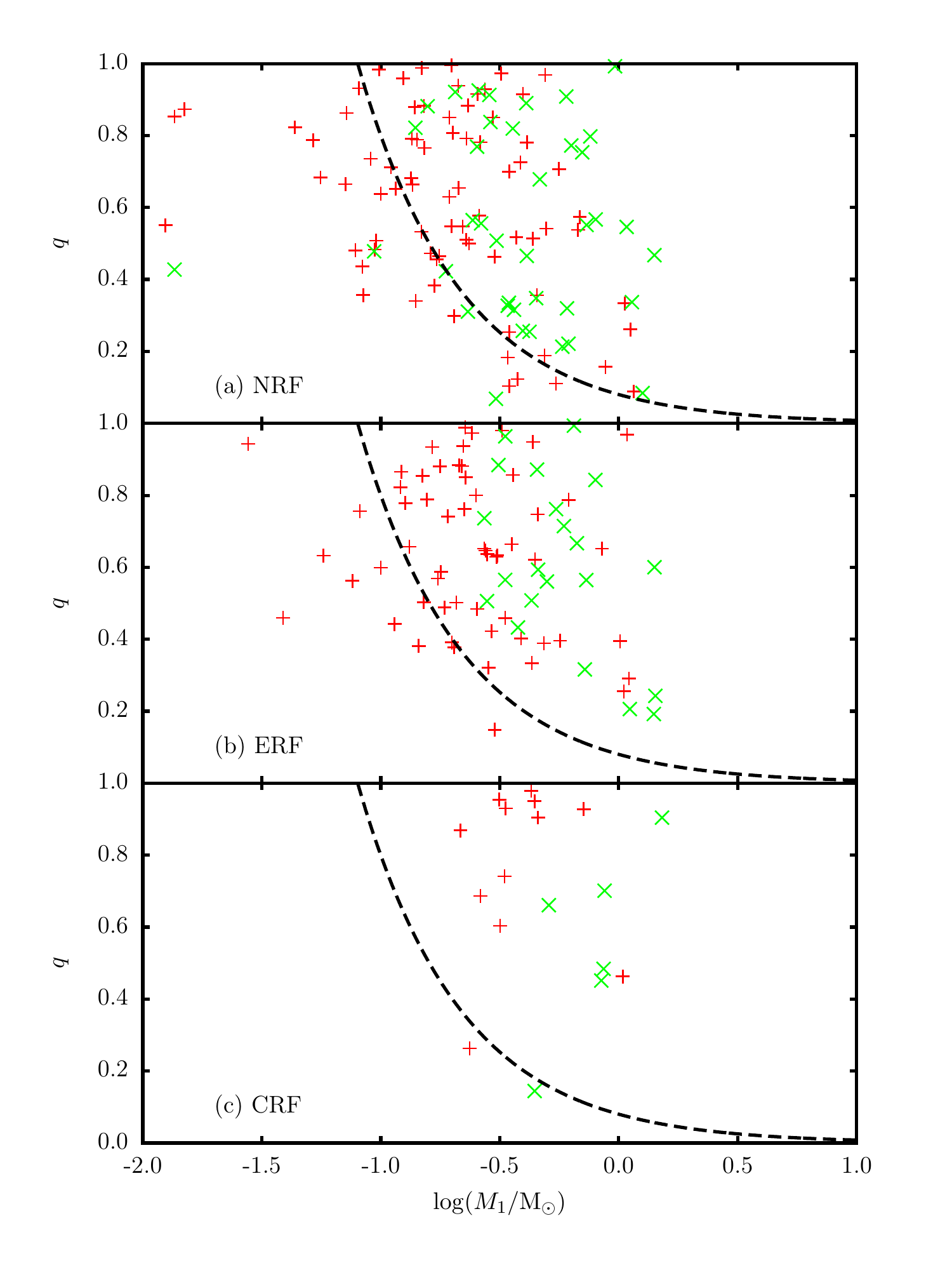}
\caption{Mass ratios plotted against primary mass, for simulations with (a) NRF, (b) ERF, and (c) CRF. A red + represents an {\sc inner} orbit. A green $\times$ represents an {\sc outer} orbit. Orbits beneath the dashed black line have brown-dwarf secondaries.}
\label{primass_massratio}
\end{figure}

\subsubsection{Intercomparison of simulations invoking different radiative-feedback prescriptions}%

Fig. \ref{semimajor} shows the distributions of orbital semimajor axis, $a$, from the simulations with (a) NRF, (b) ERF and (c) CRF; also shown are the separate distributions for {\sc inner} orbits (orbits for which both components are stars) and {\sc outer}  orbits (orbits for which at least one component is multiple, and possibly both). The arithmetic means and standard deviations of $\log_{_{10}}(a/{\rm au})$ from these distributions are given in Table \ref{mult_stats}. Modulo the fact that simulations with CRF deliver many fewer multiple systems (and only one closer than $\sim\!1\,{\rm au}$), the distributions of semi-major axis obtained with different feedback prescriptions are broadly similar. In particular, the semimajor axes delivered by the simulations span four orders of magnitude, from $0.1\,{\rm au}$ to $10^3\,{\rm au}$. 

Fig. \ref{primass_semimajor} shows the variation of orbital semi-major axis, $a$, with primary mass, $M_{_1}$, for primary masses in the range $\sim 0.03\,{\rm M}_{_\odot}$ to $\sim 1\,{\rm M}_{_\odot}$, from the simulations with (a) NRF, (b) ERF and (c) CRF. For all three sets of simulations, the semimajor axis tends to increase with primary mass. For example, the simulation results obtained with ERF can be fit approximately with ${\bar a}\!\sim\!41\,{\rm au}\,(M_{_1}/{\rm M}_{_\odot})^{1.8}$ (i.e. the solid line on Fig. \ref{primass_semimajor}b). However, there is a lot of scatter, the mass range over which this relation holds is small (about one and a half orders of magnitude), and part of the increase of $a$ with $M_{_1}$ may derive from the fact that, for the {\sc outer} orbits of hierarchical multiples, the "primary mass" is actually the sum of the masses of two or more stars.

Fig. \ref{FIG:s2a} shows, for stars that end up in multiple systems, the initial separations between the corresponding sink particles at their creation, $s$, against orbital semi-major axis at the end of the simulation, $a$, for the simulations with (a) NRF, (b) ERF and (c) CRF. The majority of pairings come into being at separations in, or close to, the range $70\,{\rm au}\la s\la 140\,{\rm au}$. We note that this is the sweet spot for disc fragmentation \citep{WS06}, where the cooling time of a proto-fragment in a massive protostellar disc is less than the dynamical timescale on which it contracts; this means that such proto-fragments are unlikely to undergo an adiabatic bounce and be sheared apart, and therefore they are likely to condense out to form additional stars. We are currently working on developing an algorithm to determine what fraction of stars are actually created by disc fragmentation in the simulations. We note that, with one exception, all pairs are created at separations exceeding $\sim\! 10\,{\rm au}$ (i.e. $\sim 50\,r_{_{\rm SINK}}$) so the dynamics of sink formation should be well resolved; close orbits are populated subsequently by scattering and orbital decay.

\subsubsection{Comparison of simulations with observation}%

{\it Young embedded stars}

\noindent The range of orbital semi-major axes generated by the simulations (Fig. \ref{semimajor}) can be compared with observation. The upper limit ($\sim 10^3\,{\rm au}$) is in good agreement with observations of young protostellar binaries in Ophiuchus \citep{RKL05}. The lower limit ($\sim 0.1\,{\rm au}\!\equiv\!20R_{_\odot}$) is set by the resolution of the simulations. Since sink particles have radii $r_{_{\rm SINK}}\!\sim\!0.2\,{\rm au}$, and their gravitational field is softened on the same scale, the formation of very close orbits is inhibited.

In the NRF and ERF simulations, the distributions of {\sc inner} semi-major axis peak at $\sim\!1.6\,\mathrm{au}$, whereas the distributions of {\sc outer} semi-major axis peak at $\sim\!80\,\mathrm{au}$. Astrometric observations of pre-Main Sequence stars can usually only resolve separations $a\gtrsim20\,\mathrm{au}$ \citep[see][and references therein]{KPPG12,KGPP12}, and from Fig. \ref{semimajor} nearly all orbits in this range from the simulations are {\sc outer} orbits of hierarchical systems. In other words, few of the {\sc inner} orbits of hierarchical systems produced in the simulations would be observationally resolvable. This suggests that many observed ``binaries'' in embedded populations might actually be partially resolved higher-order systems. The CRF simulations show a similar trend, except that the distribution of {\sc inner} semimajor axes peaks at $\sim5\,\mathrm{au}$. \\

\noindent {\it Mature field stars}

\noindent The increase in the mean orbital semi-major axis with increasing primary mass that is seen in the simulation results with ERF (Fig. \ref{primass_semimajor}b; $\left.{\bar a}\!\sim\!41\,{\rm au}\,(M_{_1}/{\rm M}_{_\odot})^{1.8}\right)$ is echoed in observations of field stars \citep{DK13}. However, in field stars the slope is significantly shallower, ${\bar a}\!\sim\!50\,{\rm au}\,(M_{_1}/{\rm M}_{_\odot})^{1.1}$ (i.e. from ${\bar a}\!\sim\!50\,{\rm au}$ at $M_{_1}\!\sim\!{\rm M}_{_\odot}$ to ${\bar a}\!\sim\!4\,{\rm au}$ at $M_{_1}\!\sim\!0.1\,{\rm M}_{_\odot}$). The principal reason for the difference is that observed very low-mass binaries typically have separations of order $4\,{\rm au}$, whereas the very-low mass binaries formed in the simulations are much closer. This could be because the gas dynamics is too dissipative, and therefore close binaries are hardened too effectively.

\subsection{Mass ratios}\label{SEC:q}%

\subsubsection{Intercomparison of simulations invoking different radiative-feedback prescriptions}%

Fig. \ref{massratios} shows the distribution of mass ratios, $q=M_{_2}/M_{_1}$ (where $M_{_1}$ and $M_{_2}$ are the masses of, respectively, the primary and the secondary), from the simulations with NRF, ERF and CRF. Also shown are the separate distributions for {\sc inner} and {\sc outer} orbits, and power-law fits of the form $p_q\!\propto\! q^{\gamma}$. If $\gamma=0$, all mass ratios are equally probable; increasing values of $\gamma$ indicate an increasing bias towards systems in which the secondary has comparable mass to the primary. Values of $\gamma$ are given in Table \ref{mult_stats}. In all cases $\gamma>0$, so there is some preference for a companion of comparable mass, irrespective of the treatment of radiative feedback. The trend is somewhat stronger for {\sc inner} orbits than {\sc outer} orbits; i.e. the components of true binary systems and the inner pairings of hierarchical multiples are more likely to be of comparable mass than the components of intermediate and outer pairings in hierarchical multiples.

Fig. \ref{primass_massratio} shows the mass ratio, $q$, plotted against primary mass, $M_{_1}$, for the simulations with NRF, ERF and CRF. This shows that, in the simulations, low mass ratios are more common in systems with high-mass primaries. Notwithstanding this trend, in all the simulations, very low-mass stars (VLMSs) and brown dwarfs (BDs) are more commonly found orbiting M-dwarf primaries,  and other VLMSs and BDs, than solar-type primaries.

\subsubsection{Comparison of simulations with observation}%

{\it Young embedded stars}

\noindent The values of $\gamma$ obtained for simulations with NRF ($\gamma\!=\!0.6\pm0.1$), ERF ($\gamma\!=\!0.9\pm0.2$) and CRF ($\gamma\!=\!1.3\pm0.5$) are all consistent with observational estimates of $0.2\!\lesssim\!\gamma\!\lesssim\!1$ for class II and class III protostars \citep{KIM11,KH12}. \\

\noindent {\it Mature field stars}

\noindent \citet{DK13} report that observational estimates of $\gamma$ for systems in the field with $a\!<\!5\,\mathrm{au}\;(>\!5\,{\rm au})$ are systematically greater (less) than for the overall population; likwise, observational estimates of $\gamma$ for {\sc inner} ({\sc outer}) orbits in higher-order multiple systems in the field are systematically greater (less) than for the overall population. This means that, in the field, stars in close orbits tend to have comparable masses, whereas stars or systems in wider orbits tend to have a broader range of mass ratios. This trend is the same as is seen in the simulations (see Fig. \ref{massratios}).

The trend found in the simulations, particularly those invoking ERF, that VLMSs and BDs are more commonly found orbiting M-dwarf primaries, and other VLMSs and BDs, than solar-type primaries, is also in agreement with field observations \citep[e.g.][]{HJ06,DK13}, which find that there is (i) a tendency for binary systems with VLMS and BD primaries to have high mass ratios ($q\!\ga\!0.5$), and (ii) a lack of BDs orbiting Sun-like stars (the Brown Dwarf Desert).

We note that the mass ratio of a multiple system may be the one property that changes little due to processing between young embedded populations and mature field populations. This is because the processes dominating the disruption of multiple systems are likely to be so impulsive that they are almost as disruptive to systems with high binding energy as systems with low binding energy \citep{PR13}.

\begin{figure}
\centering
\includegraphics[width=0.8\columnwidth]{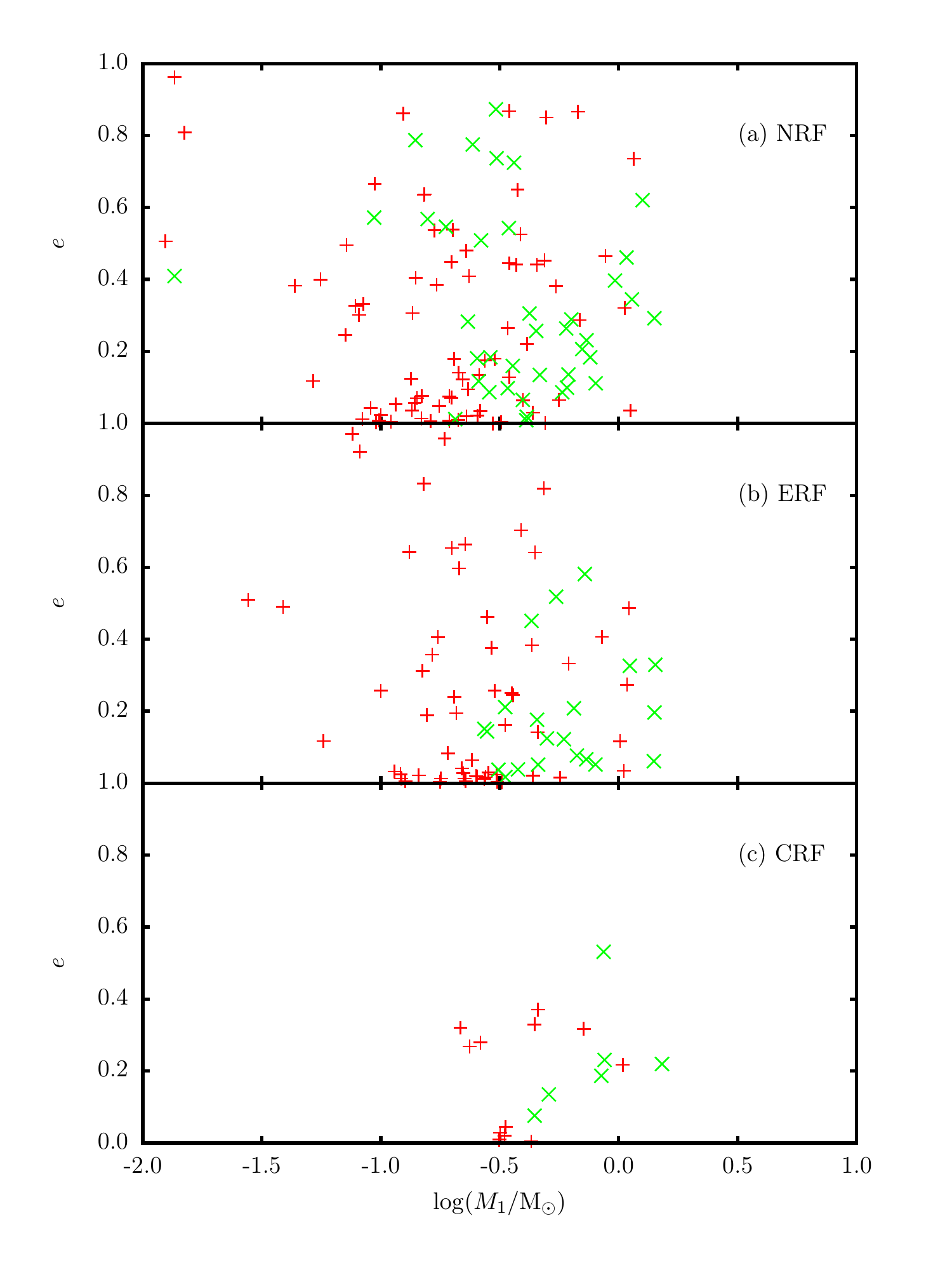}
\caption{Eccentricities plotted against primary mass, for simulations with (a) NRF, (b) ERF, and (c) CRF. A red + represents an {\sc inner} orbit. A green $\times$ represents an {\sc outer} orbit.}
\label{FIG:ECCM1}
\end{figure}

\begin{figure}
\centering
\includegraphics[width=0.8\columnwidth]{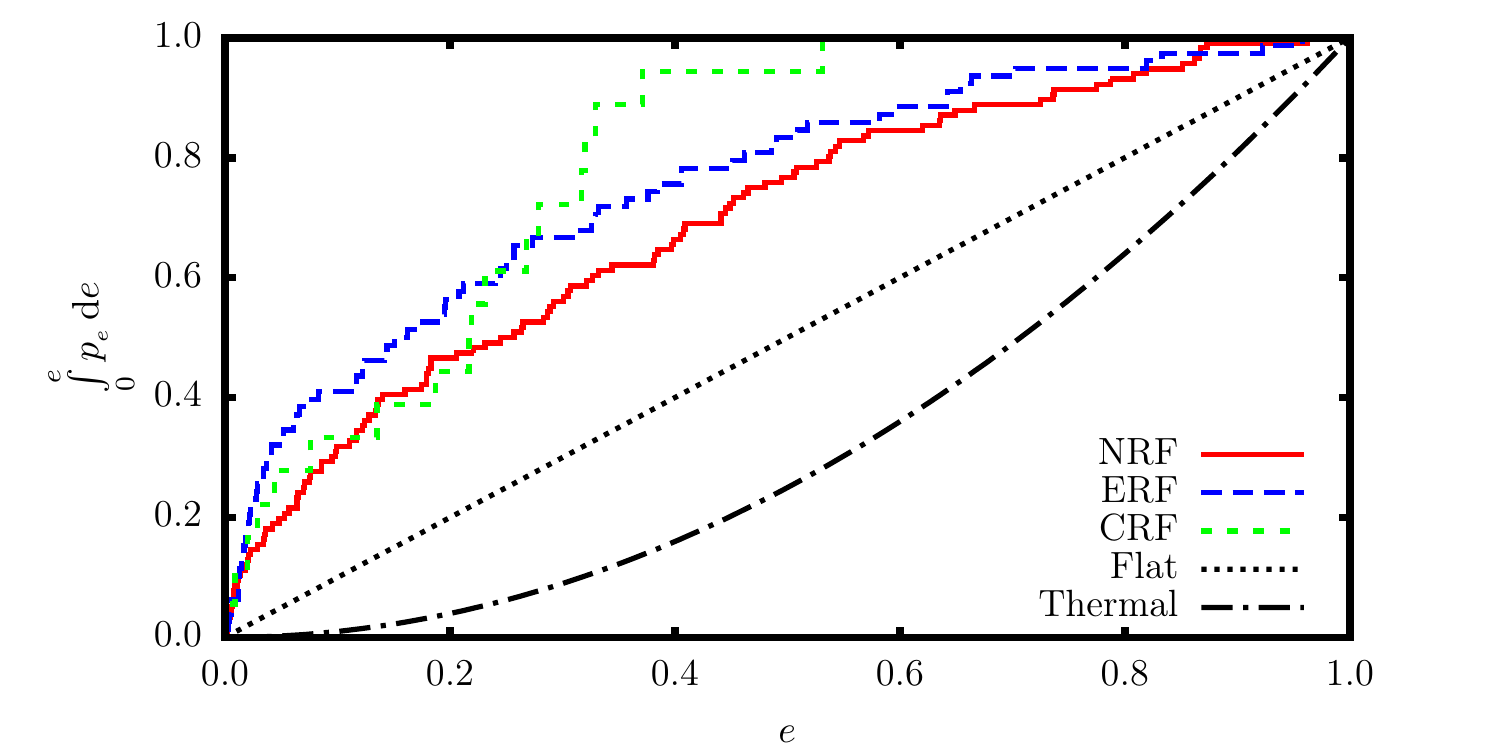}
\caption{The cummulative distribution of eccentricities, for simulations with (a) NRF, full red line; (b) ERF, dashed blue line; and (c) CRF, dashed green line. The black dotted and dot-dash lines show the cummulative distributions of eccentricity for, respectively, the uniform case (i.e. $p_e\!=\!1,\;$hence$\;\int p_ede\!=\!e$) and the thermal case (i.e. $p_e\!=\!2e,\;$hence$\;\int p_ede\!=\!e^2$).}
\label{FIG:ECCCUM}
\end{figure}

\subsection{Eccentricities}\label{SEC:ECC}%

\subsubsection{Intercomparison of simulations invoking different radiative-feedback prescriptions}%

Fig. \ref{FIG:ECCM1} shows orbital eccentricities plotted against primary mass, for the systems formed in the simulations. For the simulations with NRF and ERF, there is a broad range of eccentricites, but a preference for low eccentricities, and no significant correlation with primary mass; with CRF, only eccentricities $e\!\lesssim\!0.5$ are obtained.

Fig. \ref{FIG:ECCCUM} shows the cummulative distribution of eccentricities, for the systems formed in the simulations, irrespective of mass. The distributions obtained with NRF and ERF are very similar to one another, and show a significant preference for low eccentricities; these distributions are very different from either a flat distribution (i.e. no preferred eccentricity, $p_e\!=\!1,\;$therefore$\;\int p_ede\!=\!e$) or a thermal distribution (which favours high eccentricities, $p_e\!=\!2e,\;$therefore$\;\int p_ede\!=\!e^2$).

\subsubsection{Comparison of simulations with observation}%

{\it Mature field stars}

\noindent The observed distribution of eccentricities reported by \citet[][their Fig.3]{DK13} for M dwarfs is very similar to that obtained in the simulations with NRF and ERF (see Fig. \ref{FIG:ECCCUM}a,b), albeit with a somewhat weaker preference for low eccentricities than in the simulations. This might indicate that the gas dynamics in the simulations is too dissipative and therefore tidal circularisation is too effective.

The distribution that \citet{DK13} present for G dwarfs (Sun-like stars) is close to being flat, apart from a lack of low- and high-eccentricity systems (i.e. $0.1\la e\la 0.9$). The simulations produce too few G dwarfs to check whether the observed distribution of eccentricities is reproduced in the simulations.

\begin{figure}
\centering
\includegraphics[width=0.8\columnwidth]{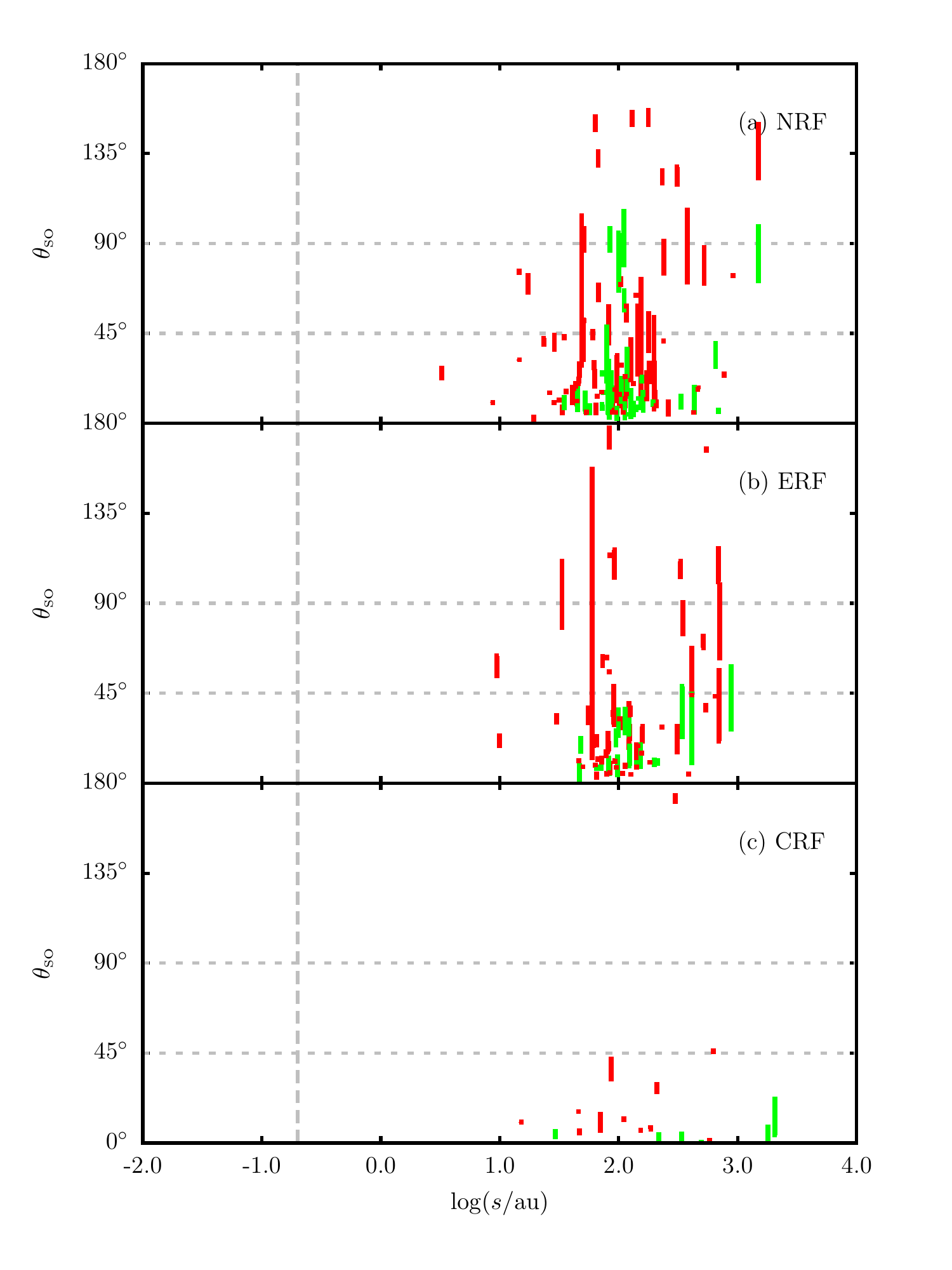}
\caption{This plot shows, for each orbit, the angles between the orbital angular momentum and the two spin angular momenta of its components, plotted against the initial separation, $s$, for systems and subsystems formed in the simulations with (a) NRF, (b) ERF, and (c) CRF. The two angles of a single orbit are connected by a vertical line; if this line is short (long), the spins are well (poorly) aligned one with another. A red line is used for {\sc inner} orbits, and a green line for {\sc outer} orbits. The horizontal dashed lines indicate angles of $45^{\rm o}$ and $90^{\rm o}$. The vertical dashed line is at the radius of a sink particle, $r_{_{\rm SINK}}\sim 0.2\,{\rm au}$.}
\label{FIG:SO_BSEP}
\end{figure}

\begin{figure}
\centering
\includegraphics[width=0.8\columnwidth]{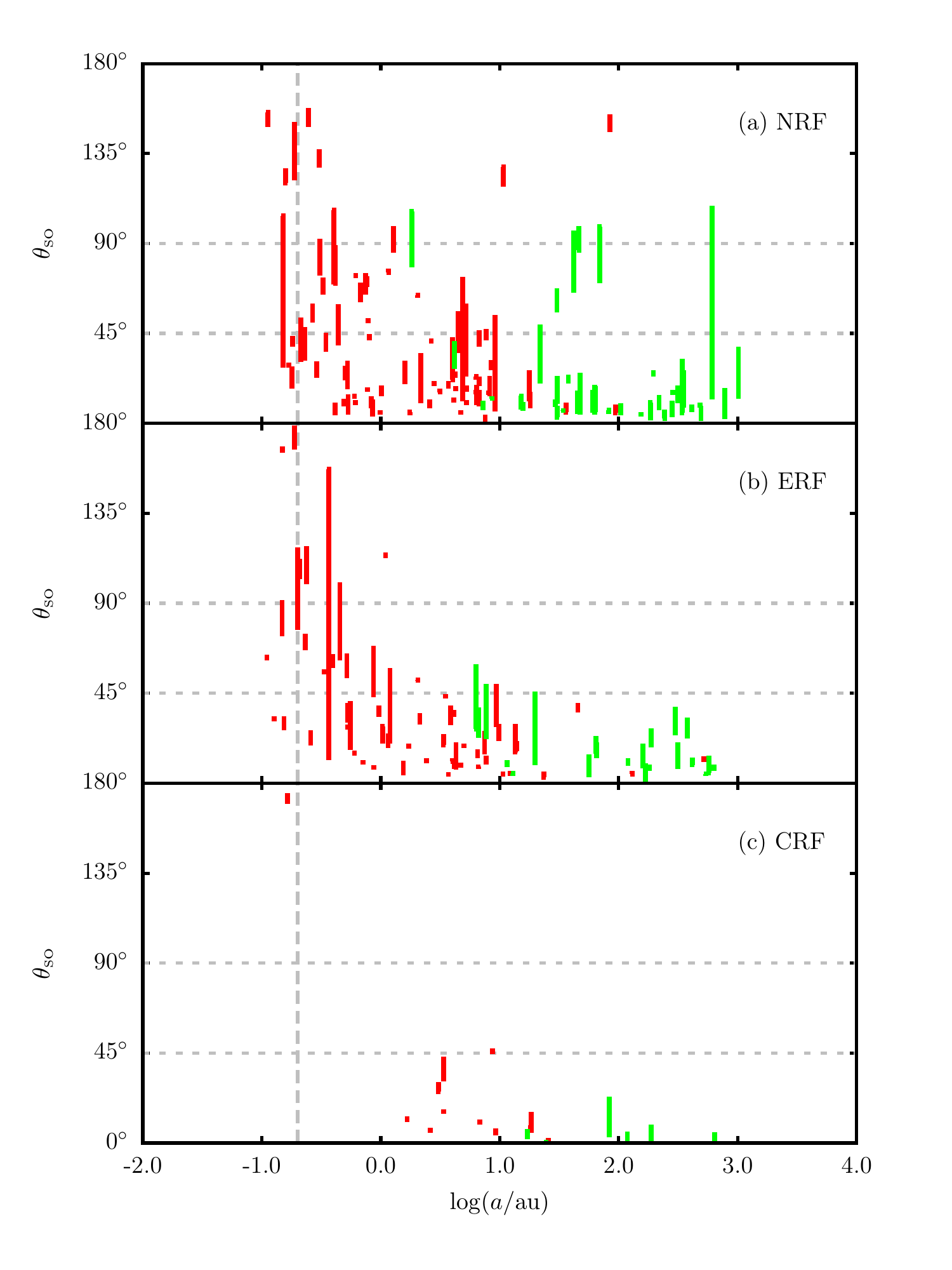}
\caption{As Fig. \ref{FIG:SO_BSEP}, but the abscissa is now the semi-major axis, $a$ at the end of the simulation, $0.2\,{\rm Myr}$.}
\label{FIG:SO_SMA}
\end{figure}

\subsection{Spin-orbit alignment}\label{SEC:SOA}%

\subsubsection{Intercomparison of simulations invoking different radiative-feedback prescriptions}%

Fig. \ref{FIG:SO_BSEP} shows, for the simulations with (a) NRF, (b) ERF and (c) CRF, and for each orbit, the two angles between the orbital angular momentum and the spin angular momenta of the two components, plotted against initial separation and connected by a thin vertical line. We distinguish {\sc inner} orbits (red) from {\sc outer} orbits (green); for the latter, the spin of a component that is a sub-system is its total angular momentum and its initial position is the centre of mass of the constituent sink particles. In most cases these angles are less than $45^{\rm o}$, especially for the outer orbits, but there are a few systems in which the spins are tilted by more than $90^{\rm o}$, relative to the orbit. In most cases the two spins associated with a given orbit are within $\sim\!10^{\rm o}$ of each other (i.e. the thin vertical connecting lines are short.

Fig. \ref{FIG:SO_SMA} shows the same angles, but now plotted against the semi-major axis, $a$, at the end of the simulation ($0.2\,{\rm Myr}$). In the simulations, once a star has formed, its spin angular momentum can only be changed by accretion; in contrast, the orbital angular momenta -- and hence also the notional spins -- of sub-systems can also be changed by tidal forces. Figs. \ref{FIG:SO_BSEP} and \ref{FIG:SO_SMA} are therefore compatible with the hypotheses that (a) many of these systems have formed by disc fragmentation, and therefore have their spins approximately aligned with their orbits \citep[to within $\sim 20^{\rm o}$; cf.][]{SW09a}, and (b) most of the {\sc inner} orbits have been populated by scattering and/or accretion of low angular momentum material, and therefore they have had their orbits and spins significantly modified.

The mean angles between orbit and spin ({\sc SO}), and spin and spin ({\sc SS}), for the different types of orbit ({\sc inner} and {\sc outer}) and the different treatments of radiative feedback (NRF, ERF and CRF) can be summarised as follows.
\begin{center}
\begin{tabular}{lllll}
 & {\sc inner} & & {\sc outer} & \\
 & $\overline{\Delta\theta}_{_{\rm SO}}$ & $\overline{\Delta\theta}_{_{\rm SS}}\hspace{1.0cm}$ & $\overline{\Delta\theta}_{_{\rm SO}}$ & $\overline{\Delta\theta}_{_{\rm SS}}$ \\
NRF\hspace{1.0cm} & $42^{\rm o}$ & $19^{\rm o}$    & $23^{\rm o}$    & $20^{\rm o}$ \\
ERF               & $41^{\rm o}$ & $14^{\rm o}$ & $17^{\rm o}$    & $19^{\rm o}$ \\
CRF               & $30^{\rm o}$ & $\;\,4^{\rm o}$ & $\;\,5^{\rm o}$ & $\;\,9^{\rm o}$ \\
\end{tabular}
\end{center}

\subsubsection{Comparison of simulations with observation}%

{\it Mature field stars}

\noindent In comparing these results with observation, we should consider the {\sc inner} and {\sc outer} orbits separately, and be mindful of two considerations. (i) The spin of a sink particle cannot always be interpreted as the spin of the star it represents, since much of the angular momentum in the sink may be attributable to the unresolved part of an attendant accretion disc within the sink radius (here $r_{_{\rm SINK}}=0.2\,{\rm au}$), rather than the star itself ($r\la0.02\,{\rm au}$). (ii) For {\sc outer} orbits, at least one component (and possibly both) is not a simple star but a subsystem in an high-order multiple.

{\sc Inner orbits.} \citet{H94} finds that close binaries in the field tend to have the intrinsic spins of their component stars approximately aligned with the orbit, but tight subsystems in higher-order multiples can be very poorly aligned. This is approximately consistent with the simulations performed with NRF and ERF (red lines on Fig. \ref{FIG:SO_SMA}), but not with CRF.

{\sc Outer orbits.} The observational data of \citet{ST02} indicate that the orbits in visual hierarchical triples in the field are only weakly aligned, i.e. $\overline{\Delta\theta}_{_{\rm SO}}\!\sim\!67^{\rm o}\pm 9^{\rm o}$, with some actually involving counter-rotation ($\Delta\theta_{_{\rm SO}}\!>\!90^{\rm o}$). Similarly, \citet{H94} finds that the intrinsic spins of stars on {\sc outer} orbits in higher-order multiples can be very poorly aligned with the orbit. This is at variance with the values of $\overline{\Delta\theta}_{_{\rm SO}}$ for {\sc outer} orbits in the simulations, which indicate a much stronger degree of alignment. One possible explanation for this is that subsequent processing knocks the wider orbits around, as simulated by \citet{ST02}.

\section{Discussion and conclusions}\label{discusion}%
 
We have analysed the statistical properties of the multiple systems formed in the simulations of Ophiuchus-like cores reported in LWH14. Cores have been created with a distribution of mass, size, projected shape, temperature and radial velocity dispersion that matches the distributions and correlations seen in Ophiuchus. Each core has been evolved for $0.2\,{\rm Myr}$ (about five freefall times), three times, once assuming no radiative feedback from the stars that form, once regulating accretion onto stars so that it is episodic and taking account of the resulting episodic radiative feedback, and once allowing continuous accretion onto stars and hence continuous radiative feedback. The statistical properties of the resulting stars, in particular those relating to multiplicity, have then been analysed to determine the statistical differences that derive from the different treatments of radiative feedback. The results have also been compared with the available observations, with a view to determining which type of radiative feedback fits the observations best. 

\subsection{Limitations}\label{SEC:LIM}%

All of our conclusions require that the initial conditions actually are realistic representations of conditions in Ophiuchus, that all the important deterministic physics has been included in the simulations {\it and} that the numerical code faithfully captures that physics. For example, it remains to be demonstrated that the prescription we have used for episodic feedback, and in particular the duty cycle that it delivers, are realistic. These are serious caveats, and should be born in mind.

In addition, the simulations lack any treatment of magnetic fields or mechanical feedback. It is not \emph{a priori} clear what effect the inclusion of magnetic fields might have. If it had a significant effect, the likelihood is that it would make the gas dynamics more dissipative and reduce the amount of fragmentation, which would probably worsen the agreement with observation. The omission of mechanical feedback means that most of the core material should accrete onto the stars, unless the initial velocity field is very divergent. \citet{MM00} estimate the effect of mechanical feedback in an isolated core, and conclude that the efficiency of star formation, $\eta$, should satisfy $0.25\la\eta\la 0.70$. At the end of the simulations ($0.2\,{\rm Myr}$), the mean efficiency (i.e. the fraction of the initial core mass that has gone into forming stars) is $\eta\!\sim\!0.7$, which is just compatible with \citet{MM00}. $\eta$ is poorly constrained by observation, since, between the time when the mass of a core is measured and the time when the core has finished forming stars, it is likely to accrete from its surroundings (e.g. along the filament in which it is embedded), and so the efficiency could notionally exceed unity; for example, \citet{HWGW13} have presented statistical arguments that favour $0.70\la\eta\la 1.3$, when ongoing accretion is allowed, so the simulations are also compatible with this estimate. 

With these caveats, we conclude that episodic radiative feedback delivers stars whose statistical properties agree best with the observations of young embedded stellar populations {\it and} mature field-star populations. We conclude by summarising these properties, and indentifying those that are not reproduced.

\subsection{Episodic radiative feedback}%

Simulations of the fragmentation of Ophiuchus-like prestellar cores, moderated by episodic accretion and radiative feedback, appear to reproduce quite well {\it both} the -- albeit as yet poorly constrained and rather limited -- multiplicity statistics observed for young embedded populations in star formation regions, {\it and} most of the the more robust statistics and relative trends seen in mature field-star populations.

Specifically, these simulations deliver an IMF, multiplicity frequencies, and distributions of (i) system order, (ii) semi-major axis and (iii) mass-ratio, which are compatible with what is observed in Ophiuchus and other {\it young embedded stellar populations}. 

They also deliver an IMF, a ratio of low-mass stars to brown dwarfs, a distribution of semi-major axes, a distribution of mass ratios --- including (a) the systematic variations between {\sc inner} orbits (which prefer components of comparable mass) and {\sc outer} orbits (which do not), (b) the Brown Dwarf Desert, and (c) the preference for companions of comparable mass in very low-mass systems --- plus a distribution of spin/orbit alignment for {\sc inner} orbits, all of which agree with observations of {\it mature field-star populations}.

The distributions of multiplicity frequency, higher-order systems, and spin/orbit alignment for {\sc outer} orbits, are only compatible with observations of mature field-star populations if -- as might be expected \citep[e.g.][]{MK11} -- dynamical processing during their dispersal into the field destroys some multiple systems, in particular the outer orbits of high-order multiples, and perturbs the surviving orbits. 

However, there are two specific concerns with the simulations. First, they produce low-mass binaries whose orbits are too small. Second, although the overall distribution of eccentricities mimics quite well that seen in field M Dwarfs, there is a somewhat larger excess of low eccentricities in the simulations. Both of these concerns suggest that the gas dynamics in the simulations is too dissipative, particularly on small scales. A similar concern has been reported by other workers in this field \citep[e.g.][]{B14}.

We also note the following. (i) The simulations suggest that in some observed wide binary systems the primary may be an unresolved close binary. (ii) Many of the stars appear to be formed by disc fragmentation at radii $70\,{\rm au}\la R\la 140\,{\rm au}$, i.e. the sweet spot for disc fragmentation identified by \citet{WS06}.

\subsection{No radiative feedback}%

Simulations of the same cores, but with no radiative feedback, produce rather similar statistics to those with episodic radiative feedback, except that the simulations produce too many brown dwarfs, so the IMF is skewed to low masses and the ratio of low-mass stars to brown dwarfs is consequently too low. We are not suggesting that there might actually be no radiative feedback; these simulations are simply included as a point of reference.

\subsection{Continuous radiative feedback}%

Simulations with continuous accretion and radiative feedback appear not to work. There are far too few brown dwarfs, so the IMF is skewed to high masses, and the ratio of low-mass stars to brown dwarfs is too large. The multiplicity frequency is too low, there are too few close orbits, and there is no tendency for the semi-major axis to increase with primary mass. The distribution of orbital eccentricities is wrong (all $e\la 0.5$), and the spins and orbits are too well aligned.

\subsection{Summary}%

Modulo the limitations discussed in \S \ref{SEC:LIM}, our simulations suggest that the star formation observed in Ophiuchus can only be representative of global star formation if accretion onto -- and hence radiative feedback from -- young stars is episodic.

\section*{Acknowledgements}

OL and APW gratefully acknowledge the support of a consolidated grant (ST/K00926/1) from the UK STFC. SW acknowledges support from the DFG priority programme 1573, "Physics of the ISM", and the Bonn-Cologne Graduate School of Physics \& Astronomy.

\bibliographystyle{mn2e}
\bibliography{refs}

\appendix

\section{Estimating the population multiplicity frequency}
\label{SEC:MF_APP}

The population multiplicity frequency $mf$ can be estimated for a sample of steller systems with $S$ singles and $M$ multiples by invoking Bayes' theorem:
\begin{equation}
  P(mf|S,M)=\frac{P(S,M|mf)\,P(mf)}{P(S,M)}\,.
  \label{bayes}
\end{equation}
Here, $P(mf|S,M)$ is the posterior probability of $mf$ given $S$ and $M$. $P(S,M|mf)$ is the likelihood of sampling $S$ singles and $M$ multiples given $mf$, i.e. the binomial distribution,
\begin{equation}
  P(S,M|mf)=\frac{(S+M)!}{S!\,M!}\,mf^M\,(1-mf)^S\,.
\end{equation}
$P(mf)$ is the prior probability of $mf$. As we have no \emph{a priori} evidence, we give all values of $mf$ equal probability, i.e. $P(mf)\!=\!1$\,. The denominator of Eqn. (\ref{bayes}) is the integral of the numerator over all vaules of $mf$, i.e.,
\begin{equation}
  P(S,M)=\int\limits_{0}^{1}P(S,M|mf)\,P(mf)\,\mathrm{d}mf\,.
\end{equation}
The posterior distribution of $mf$ is therefore
\begin{equation}
  P(mf|S,M)=\frac{mf^M\,(1-mf)^S}{\int_0^1mf^M\,(1-mf)^S\,\mathrm{d}mf}\,,
\end{equation}
which is the beta distribution.

An estimate of $mf$ can be obtained by calculating the mean $\mu_{mf}$ and standard deviation $\sigma_{mf}$ of $P(mf|S,M)$:
\begin{equation}
  \begin{split}
    \mu_{mf}&=\int\limits_{0}^{1}mf\,P(mf|S,M)\,\mathrm{d}mf\,,\\
    \sigma_{mf}^2&=\int\limits_{0}^{1}(mf-\mu_{mf})^2\,P(mf|S,M)\,\mathrm{d}mf\,.
  \end{split}
\end{equation}
Note that as $\mu_{mf}$ is the mean of a probability density function, it will always be greater than zero (less than one) even if the entire sample is single (multiple).

\label{lastpage}
\end{document}